\documentclass[trackchanges]{aastex701}

\usepackage[utf8]{inputenc}
\usepackage{overpic}
\usepackage{multirow} 
\usepackage{arydshln}  
\usepackage{booktabs}  
\usepackage{hyperref}
\usepackage{CJK}
\graphicspath{{./}{figures/}}
\graphicspath{{./}{./_fig_2604/}}

\begin{document}

\title{A GeV Gamma-Ray Spectral Excess in the Fermi Blazar 4FGL J0136.5$+$3906}

\author[orcid=0000-0002-9071-5469,gname='Shi-Ju',sname='Kang']{Shi-Ju Kang}
\affiliation{School of Physics and Electrical Engineering, Liupanshui Normal University, Liupanshui, Guizhou, 553004, People's Republic of China}
\email[show]{kangshiju@alumni.hust.edu.cn} 
\email[]{kangshiju@lpssy.edu.cn}

\author[orcid=0009-0008-5422-7485,gname='Yue',sname='Yin']{Yue Yin}
\affiliation{School of Physics and Electrical Engineering, Liupanshui Normal University, Liupanshui, Guizhou, 553004, People's Republic of China}
\email[]{yueyin@lpssy.edu.cn}

\author[orcid=0000-0003-0170-9065,gname='Yong-Gang',sname='Zheng']{Yong-Gang Zheng}
\affiliation{Department of Physics, Yunnan Normal University, Kunming, Yunnan, 650092, People's Republic of China}
 \email[show]{ynzyg@ynu.edu.cn}

\author[orcid=0000-0003-4773-4987,gname='Qingwen',sname='Wu']{Qingwen Wu}
\affiliation{Department of Astronomy, School of Physics, Huazhong University of Science and Technology, Wuhan, Hubei, 430074, People's Republic of China}
\email[show]{qwwu@hust.edu.cn}

\author[orcid=0000-0002-5929-0968,gname='Junhui',sname='Fan']{Junhui Fan}
\affiliation{Center for Astrophysics, Guangzhou University, Guangzhou 510006, People's Republic of China}
\email[show]{fjh@gzhu.edu.cn}

\begin{abstract}
We report the detection of a $\sim${{19}} GeV gamma-ray spectral excess (peak energy in observer frame) at $\sim$5.7$\sigma$ significance (3 degrees of freedom, without correction for trials and the look-elsewhere effect) in the Fermi-LAT data of the blazar 4FGL J0136.5$+$3906, based on $\sim${17} years of observations. The excess manifests as a distinct bump-like feature in the 100 MeV$-$1 TeV spectrum, with a flux of $\sim$3.6$\rm \times10^{-10}~ph~cm^{-2}~s^{-1}$. No similar feature is detected in nearby sources or background regions. The origin of this signal cannot be explained by standard blazar emission models and remains unknown. This Letter presents the observational finding without providing a definitive physical interpretation.
\end{abstract}


\section{Introduction} 

Blazars are a subclass of active galactic nuclei with a relativistic jet oriented close to our line of sight \citep{1995PASP..107..803U}. Their broadband spectral energy distributions (SEDs) display a characteristic double-hump structure: the low-energy hump, from radio to X-rays, is attributed to synchrotron radiation from relativistic electrons in the jet. The origin of the high-energy hump, peaking in the MeV$-$TeV range, is still debated: leptonic models invoke inverse Compton scattering \citep{1992ApJ...397L...5M,1994ApJ...421..153S}, while hadronic models involve proton-synchrotron or photo-pion processes \citep{2003APh....18..593M,2013ApJ...768...54B}. For recent reviews, see \cite{2019Galax...7...20B} and \cite{2019NewAR..8701541H}.

In the Fermi-LAT energy band ($\sim100$MeV$-$1TeV), the $\gamma$-ray spectra of most blazars are well described by smooth functions such as a single power-law or a log-parabola \citep{2010ApJ...710.1271A,2015ApJ...810...14A,2022ApJS..260...53A,2023arXiv230712546B,2026arXiv260222148B}. However, with the accumulation of more than a decade of LAT data, increasingly complex spectral features have been uncovered. Notable examples include spectral breaks (a steepening at $\sim2$\,GeV in 3C\,454.3; \citealt{2009ApJ...699..817A}), spectral hardening (a rising $\nu F_{\nu}$ spectrum above $\sim10$\,GeV in M\,87 and Pictor\,A; \citealt{2019A&A...623A...2A,2025ApJ...988..268L}, or in blazar S5\,1027+74; \citealt{2025ApJ...991L...8P} and 1ES 0502+675; \citealt{2022MNRAS.511..938Z}), which is extremely rare with an occurrence rate below 0.1\% \citep{2025A&A...703A.162D}, and narrow line-like features (a $\sim7$\,GeV excess in B0516-621, possibly from axion-like particles; \citealt{2021JCAP...08..007Z}, and a tentative $\sim31$\,GeV excess in another blazar, \citealt{2026arXiv260928303K}). Most recently, \cite{2026arXiv260400579K} reported a persistent $\sim1.5$\,GeV spectral line-like anomaly in three AGNs and a tentative double-excess signature in a single blazar object reported in \cite{2026arXiv260623488K}. These findings highlight that GeV spectra can deviate from simple continua, providing potential signatures of new physics or unrecognized jet processes.

In this Letter, we report the detection of a different spectral feature: a prominent and broad excess at $\sim${{19}} \,GeV in the Fermi-LAT spectrum of the bright blazar 4FGL J0136.5+3906. Using {17} years of Pass 8 data, we find a clear bump-like feature with a local significance of $\sim$5.7$\sigma$. Unlike narrow line-like anomalies reported in other AGNs, this feature  is broad ($\sigma_{\rm signal}\sim$10 GeV), indicating a different physical origin. The feature cannot naturally be described by standard one-zone blazar jet models, nor by any known smooth continuum representation. Below we present the observational result, leaving a definitive physical interpretation for future work.

\section{Data and Method}

\subsection{Fermi-LAT Data Analysis}

We analyzed $\sim${17} years of Pass 8 (P8R3) Fermi-LAT data from 2008 August 4 to 2025 August 4, targeting the blazar 4FGL J0136.5+3906. The data reduction and analysis were performed using Fermipy v1.2.0 \citep{2017ICRC...35..824W} and the Fermi Science Tools v2.2.0. Events within a circular region of interest (ROI) of radius $15^\circ$ centered on the source were selected in the energy range from 100\,MeV to 1\,TeV. Standard quality cuts (\texttt{(DATA\_QUAL>0) \&\& (LAT\_CONFIG==1)}) were applied, and photons with zenith angles greater than $90^\circ$ were excluded to reduce contamination from the Earth's limb. The instrument response functions P8R3\_SOURCE\_V3 were used throughout.

The background model included all sources from the Fermi-LAT Fourth Source Catalog (4FGL-DR4; \citealt{2022ApJS..260...53A,2023arXiv230712546B}) within the ROI, together with the Galactic diffuse emission model (\texttt{gll\_iem\_v07.fits}) and the isotropic diffuse component (\texttt{iso\_P8R3\_SOURCE\_V3\_v1.txt}). The normalization parameters of the two diffuse components and the spectral parameters of all sources within $5^\circ$ of the target were left free during the likelihood fitting. The SED points with a test statistic TS $\ge 25$ were retained; otherwise, $95\%$ confidence level upper limits were derived.

\subsection{Spectral Model and Significance Analysis}

To search for a possible broad excess superimposed on the continuum, we performed a detailed spectral fit comparing two models:
\textbf{Null hypothesis} – a single log-parabola (LP) function:
\[
\frac{dN}{dE} = K \left(\frac{E}{E_0}\right)^{-\alpha - \beta \ln(E/E_0)},
\]
where $\alpha$ is the spectral slope at the pivot energy $E_0$, $\beta$ is the curvature parameter, and $K$ is the normalization.
\textbf{Signal hypothesis} – a log-parabola plus a Gaussian component to model a possible broad/line-like excess:
\[
N_{\mathrm{signal}}(E) = \frac{N_0}{\sigma_{\mathrm{signal}}\sqrt{2\pi}} \exp\!\left(-\frac{(E-E_{\mathrm{signal}})^2}{2\sigma_{\mathrm{signal}}^2}\right),
\]
where $N_0$ is the amplitude, $E_{\mathrm{signal}}$ the central energy, and $\sigma_{\mathrm{signal}}$ the width (standard deviation). The Gaussian width was allowed to vary freely to account for either an intrinsically narrow line or a moderately extended feature.
All three Gaussian parameters were kept free and scanned over the entire energy spectrum from 100\,MeV to 1\,TeV. The continuum parameters ($\alpha,\beta,K$) were also free in both hypotheses.

Using the \texttt{pyLikelihood} module, we first fitted the null model to obtain the maximum log-likelihood value $\mathcal{L}_{\mathrm{null}}$. We then fitted the signal model to obtain $\mathcal{L}_{\mathrm{signal}}$. The test statistic (TS) for the presence of an additional component is defined as: $\mathrm{TS} = 2(\mathcal{L}_{\mathrm{signal}} - \mathcal{L}_{\mathrm{null}})$. Under the null hypothesis, TS approximately follows a $\chi^2$ distribution with three degrees of freedom (the three additional Gaussian parameters). The local significance $\sigma$ of the excess is then derived as $\sigma = \sqrt{\mathrm{TS,n}}$ for three degrees of freedom (n), assuming Wilks' theorem.
 
To check robustness, we also performed the analysis using alternative background models (e.g., varying the normalization of the Galactic diffuse emission by $\pm10\%$), different event type selections (e.g., PSF0$-$PSF3,  EDISP0$-$EDISP3, and FRONT+BACK) and certain systematic uncertainties test. The results remained stable within the quoted uncertainties (see Appendix Table~\ref{tab2}).

\section{Results}

4FGL J0136.5+3906 is located at RA = 24.1423, Dec = 39.1 (J2000). It is a BL Lac type blazar reported in the 4FGL catalog \citep{2020ApJS..247...33A,2022ApJS..260...53A,2023arXiv230712546B}, associated with the source B3 0133+388. No spectroscopic redshift has been reported in the literature for this source; only a machine learning prediction yields a value of $z \simeq 0.18$ \citep{2022ApJS..259...55N}. Based on its synchrotron peak frequency $\nu_{\mathrm{syn}} = 1.11\times10^{16}\,\mathrm{Hz}$ \citep{2023arXiv230712546B}, it is classified as a high-synchrotron-peaked (HSP) blazar.

\begin{figure}[tp!]
\centering
\includegraphics[width=0.48\linewidth]{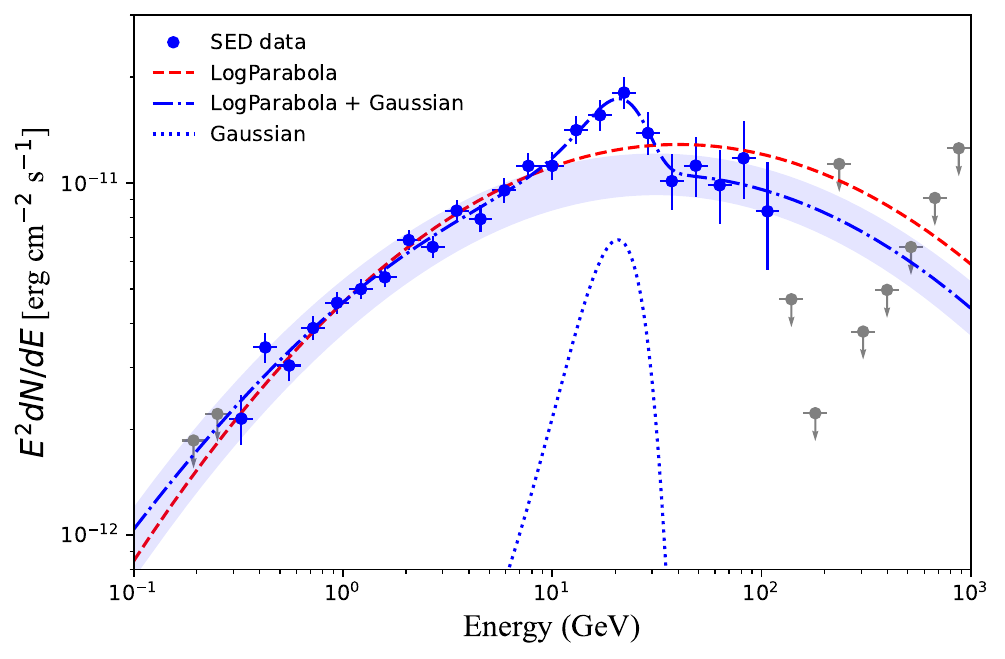}
\includegraphics[width=0.48\linewidth]{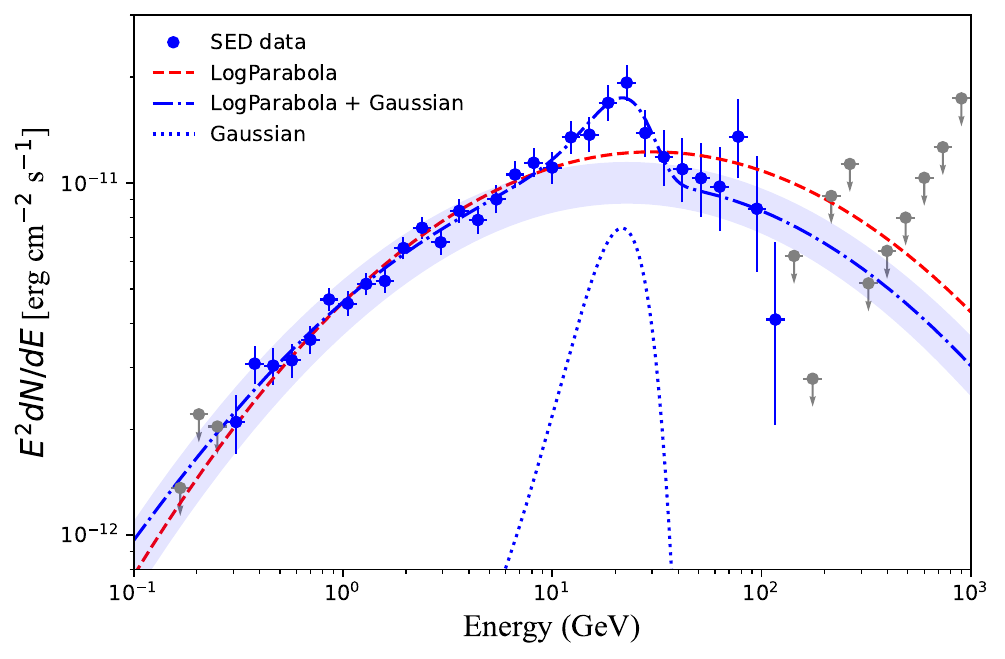}
\caption{Spectral energy distribution (SED) and spectral excess signal of 4FGL J0136.5+3906 in the 100 MeV to 1 TeV range, using {17} years of data (left panel: 35 bins, right panel: 45 bins). The SED is fitted with a pure log-parabola model (null hypothesis, red dashed line) and with a log-parabola plus Gaussian model (signal hypothesis, blue dotted-dashed line). Blue points denote SED bins with TS $> 25$; gray downward arrows indicate upper limits. The blue dotted line represents the Gaussian excess component.}
\label{fig_sed}
\end{figure}

\begin{figure}[htbp!]
\centering
\includegraphics[width=0.46\linewidth]{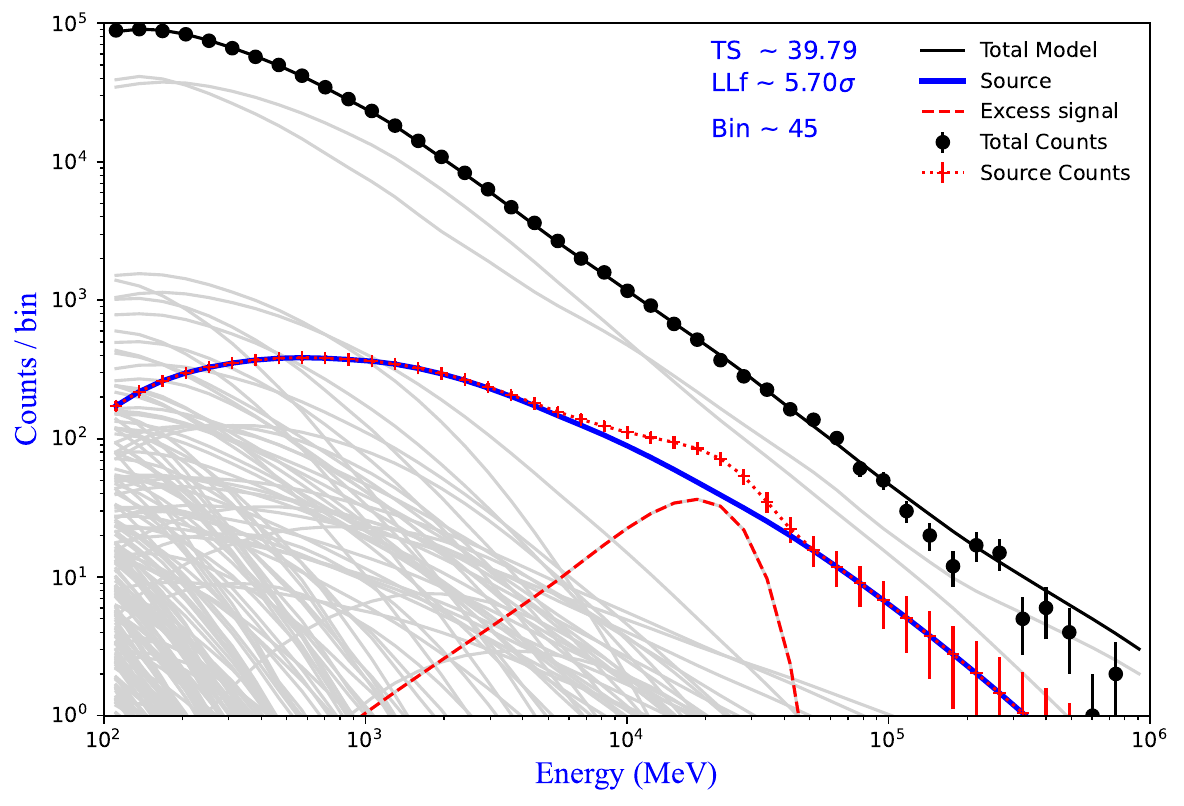}
\includegraphics[width=0.46\linewidth]{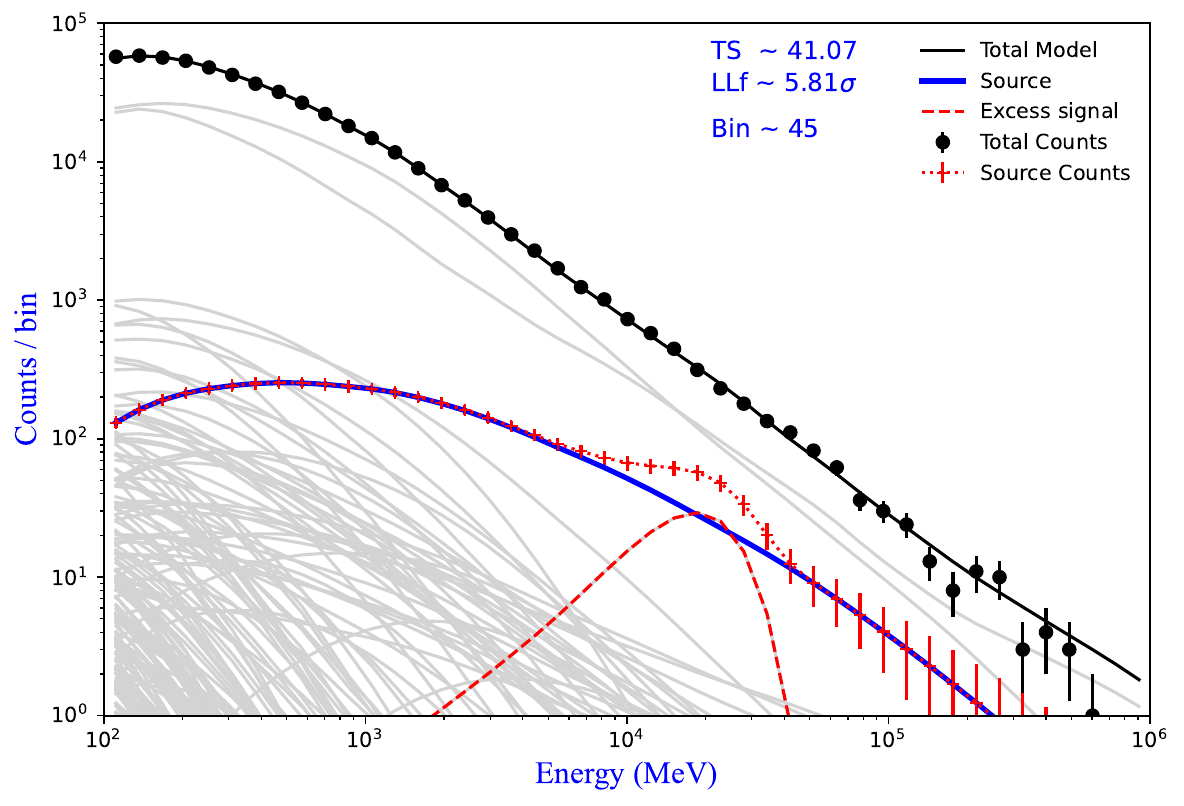}
\caption{Model-fitted count spectra of 4FGL J0136.5+3906 over {17} years (left) and a 11-year subset (right) with 45 logarithmically spaced energy bins. Black dots and black line: data and total model; light grey lines: background sources; red dot-dashed: source model; blue solid: log-parabola continuum; red dashed: Gaussian excess.}
\label{fig_count}
\end{figure}

Following standard binned maximum likelihood analysis\footnote{\url{https://fermi.gsfc.nasa.gov/ssc/data/analysis/scitools/binned_likelihood_tutorial.html}}, we constructed the time-averaged SED of 4FGL J0136.5$+$3906 using $\sim${17} years of Fermi-LAT data with 45 logarithmically spaced energy bins from 100 MeV to 1 TeV. The SED is generally well described by a single log-parabola model. However, we notice a prominent residual excess centered at $\sim${{19}} \,GeV (see Figure~\ref{fig_sed}), superimposed on a smooth log-parabolic continuum. A single log-parabola (red dashed line) describes the continuum reasonably well but leaves clear positive residuals around 19\,GeV. The alternative model (log-parabola plus Gaussian) provides a significantly better fit (blue dotted-dashed line).

To evaluate the significance of this excess, we performed a likelihood analysis comparing the pure log-parabola model (null hypothesis) with a model that incorporates an additional Gaussian component (signal hypothesis). The parameters of the Gaussian component—amplitude, central energy, and width—were allowed to vary freely and scanned across the entire energy range (see \citealt{2026arXiv260400579K}).

For the full {17}-year dataset with 45 logarithmically spaced energy bins, the best-fit Gaussian component is centered at \(E_{\mathrm{signal}} = 13.07 \pm 2.25\)\,GeV, with a width of \(\sigma_{\mathrm{signal}} = 10.00 \pm 0.02\)\,GeV in linear energy space (see Appendix Table~\ref{tab1}). This width is significantly broader than the Fermi-LAT energy resolution at these energies, indicating that the excess represents a broad superstructure rather than a narrow line. The corresponding peak of the excess in logarithmic energy space is \(E_{\mathrm{peak}} = 19.26 \pm 1.52\)\,GeV, and the integrated flux is \(F_{\mathrm{signal}} = (3.62 \pm 0.72)\times10^{-10}\)\,ph\,cm\(^{-2}\)\,s\(^{-1}\).

The maximum log-likelihood values yield a test statistic \(\mathrm{TS} = 2(\mathcal{L}_{\mathrm{signal}}-\mathcal{L}_{\mathrm{null}}) = 39.79\). Under the null hypothesis, TS approximately follows a \(\chi^2\) distribution with three degrees of freedom (the three additional Gaussian parameters). This results in a local significance of $\sim$5.7$\sigma$ (corresponding to \(\sqrt{\mathrm{TS}}\) for three degrees of freedom, where without correction for trials and the look-elsewhere effect), which surpasses the discovery threshold in high-energy astrophysics.
Even under the most conservative look-elsewhere effect correction (assuming $N_{\rm trial}$ = 500), the global significance \citep{2013PhRvD..88h2002A} declines to $\sim$4.4$\sigma$, yet the excess remains a statistically robust feature.

To assess the robustness of the $\sim$19 GeV excess against systematic uncertainties, we performed several checks (see Appendix Table~\ref{tab2}) following the procedures established in previous Fermi-LAT spectral searches (e.g., \citealt{2014JCAP...10..023A,2015PhRvD..91l2002A}). Varying the Galactic diffuse normalization by $\pm10\%$ left the significance unchanged at $\sim$5.8$\sigma$. Convolving the signal model with the LAT energy dispersion matrix produced negligible changes in the best-fit flux and central energy, confirming that energy dispersion does not artificially create or distort the excess. We also searched for the excess in two control regions --- the Earth limb and off-source fields at $|b|>30^\circ$ with no known bright gamma-ray sources --- and found no significant feature near 19 GeV (maximum TS $<7$, corresponding to a local significance $<2\sigma$), arguing against an instrumental origin. In addition to the two conversion event types (front and back) and the PSF event type (PSF0$-$PSF3) selection, we also performed a joint likelihood analysis using all four EDISP event types (EDISP0$-$EDISP3), which yielded best-fit parameters consistent with the standard analysis, indicating that energy dispersion does not significantly affect the excess. 
And we also assessed the effective area systematic uncertainty using the bracketing IRF approach \citep{2012ApJS..203....4A}, and found it to be subdominant relative to the statistical uncertainty.
The combined systematic uncertainty from these checks is comparable to the statistical uncertainty but does not affect the presence or significance of the feature. It is suggested that the excess is robust against the dominant known systematics.

Moreover, the observed excess persists when employing alternative diffuse background models, applying different event selection criteria, or adjusting the energy binning scheme (see Appendix Tables~\ref{tab1} and~\ref{tab2}). Results for other time intervals and bin numbers are consistent and are provided in the Appendix. No comparable spectral feature is detected in data from nearby sources or in background regions devoid of the source. Through PSF-based source decomposition and systematic background model scans, we further rule out the possibility of an unresolved neighboring source or a local structure in the diffuse emission, indicating that the $\sim$19 GeV excess is intrinsic to 4FGL J0136.5$+$3906 rather than due to source confusion or diffuse emission artifacts.

However, an analysis of the temporal behavior of this spectral excess --- by dividing the full 17-year span into two or four equal sub-intervals --- reveals significant flux variability. For example, during the first 8.5 years (2008.8-2017.2), the flux was approximately \(4.59 \times 10^{-10}~\text{ph}~\text{cm}^{-2}~\text{s}^{-1}\), while in the subsequent 8.5 years (2017.2-2025.8) it declined to about \(2.67 \times 10^{-10}~\text{ph}~\text{cm}^{-2}~\text{s}^{-1}\), roughly half the earlier value. Moreover, the significance of the excess decreased from TS $\sim$ 34.81 (5.27$\sigma$) to $\sim$9.90 (2.34$\sigma$). A further subdivision into four intervals showed that both the flux and the significance peaked distinctly in the second sub-period, suggesting a possible transient origin. These results imply that the excess is not sufficiently stable (see Appendix Table \ref{tab1} and Figure \ref{Fig_count_A2}). Consequently, the nature of this spectral excess feature requires more comprehensive investigation.

\section{Discussion and Conclusion}

The detection of a broad, $\sim$5.7$\sigma$ spectral excess (even without correction for trials and the look-elsewhere effect) centered at $\sim$19\,GeV with width $\sigma_{\rm signal}\sim8-10$\,GeV (see Appendix Table \ref{tab1}) in the blazar 4FGL J0136.5+3906 presents a clear challenge to conventional jet emission models \citep{2019Galax...7...20B,2019ARA&A..57..467B,2019NewAR..8701541H}. While the baseline log-parabola continuum describes the spectrum below $\sim5$\,GeV and above $\sim50$\,GeV well, it fails to account for the prominent bump in the 10--30\,GeV range that differs from the excess of PKS 1510-089 \citep{2022MNRAS.515.5242D} characterized by a broader framework (hardening). Standard one-zone leptonic or hadronic models cannot naturally produce such a broad, isolated feature without fine-tuned parameters that would overproduce neutrinos or low-energy emission not observed in this source.

In addition, although 4FGL J0136.5+3906 is an HSP BL Lac, an EC origin for the $\sim$19  GeV excess cannot be ruled out. Weak emission lines have been suggested in some BL Lacs, and EC scattering may still contribute even in low-power BL Lacs \citep{2008ApJ...684L..73A,2011ApJ...726...43A,2014ApJ...783...83L,2016MNRAS.461.1862K,2025MNRAS.544.1926C,2026ApJ..1002..199Z}. However, given the lack of direct evidence for a prominent external photon field in this source, the EC scenario remains speculative, and the origin of the excess remains inconclusive.

Several exotic scenarios have been considered, though none is compelling.
\begin{itemize}
    \item \textbf{Millisecond pulsars (MSPs)} in the host galaxy could collectively produce a gamma-ray bump \citep{2022PhRvD.105j3023C}, but would require an implausibly large number of MSPs and a more extended spatial distribution than the point-like excess.

    \item \textbf{Axion-like particles (ALPs)} oscillating in the jet's magnetic field could modulate the gamma-ray transmission \citep{2012PhRvD..86h5036T}. Yet this mechanism usually predicts oscillatory spectral features or multiple bumps \citep{2021JCAP...08..007Z}, whereas we observe only one smooth excess; the unknown magnetic field in the jet also limits the reliability of this interpretation. 

    \item \textbf{Dark matter (DM) annihilation} into $b\bar{b}$ can yield a broad gamma-ray peak. For the observer-frame peak at $\sim$19\,GeV, the rest-frame energy is $\sim$22.4\,GeV (assuming $z\approx0.18$), implying a DM mass of $\sim$90\,GeV \citep{2011PhRvD..83h3507C,2015JCAP...05..011A}. 
    However, the excess signal does not exhibit the required temporal stability, and the J-factor for a single blazar remains highly uncertain; moreover, the annihilation cross-section limits derived from the Galactic Center \citep{2025arXiv251212176W} disfavor (argue against) a dark matter annihilation origin.
    In addition, the relationship between the 19 GeV signal observed from this blazar and its potential similarity to the GeV excess detected at the Galactic center  \citep{2009arXiv0910.2998G,2011PhLB..697..412H,2024PhRvD.109l3017Z}, as well as the 20 GeV halo-like excess in the Galactic diffuse radiation \citep{2025JCAP...11..080T}, necessitates further investigation.    
\end{itemize}

Given the lack of a compelling conventional astrophysical explanation and the robustness of the signal (stable against systematic uncertainties), the origin of the $\sim$19\,GeV bump remains inconclusive. The high statistical significance ($\sim$5.7$\sigma$) firmly establishes the excess as a real feature. Distinguishing among the proposed scenarios will require coordinated multi-wavelength and multi-messenger observations, including a search for neutrinos (IceCube, KM3NeT), deeper X-ray/radio imaging to probe a possible pulsar population, and improved gamma-ray angular resolution with CTA. Until such evidence emerges, the excess stands as an intriguing observational puzzle that challenges our understanding of high-energy emission from active galactic nuclei.

Furthermore, a critical open question is whether the spectral bump observed in 4FGL J0136.5+3906 is unique to this source or represents a more general phenomenon among HSP blazars. A systematic search for similar features in a larger sample is necessary to resolve this issue, but such an investigation lies beyond the scope of the present Letter. We therefore leave this as a priority for future work.

\begin{acknowledgments}
We thank J. Li and Z. Q. Shen for valuable discussions. This research has made use of data obtained through the High Energy Astrophysics Science Archive Research Center (HEASARC) Online Service (Fermi Science Support Center) provided by NASA Goddard Space Flight Center. This work is partially supported by the National Natural Science Foundation of China (Grant No. 12163002, 12363002, 12533005, 12233007, 12433004), the National Key Research and Development Program of China (Grant No.2025YFA1614102), and the Liupanshui Science and Technology Development Project (Grant No. 52020-2024-PT-05) and the Discipline-Team of Liupanshui Normal University (LPSSY2025XKTD07).
\end{acknowledgments}

\begin{contribution}
S.-J. Kang initiated and designed the study, and performed the data analysis together with Y. Yin. 
The manuscript was drafted by S.-J. Kang and revised by Y.-G. Zheng, Q. Wu and J.-H. Fan. 
All authors discussed the results, contributed to the interpretation, approved the final version.
\end{contribution}

\facilities{Fermi(LAT)}

\software{Fermipy \citep{2017ICRC...35..824W}.}

\clearpage

\renewcommand{\figurename}{Appendix Figure}
\renewcommand{\tablename}{Appendix Table}
\renewcommand{\thefigure}{A\arabic{figure}}
\renewcommand{\thetable}{A\arabic{table}}
\renewcommand{\theequation}{A\arabic{equation}}
\renewcommand{\thepage}{A\arabic{page}}
\setcounter{figure}{0}
\setcounter{table}{0}
\setcounter{equation}{0}
\setcounter{page}{1} 

\appendix
\section{\bf Additional Figures and Tables}

\begin{figure}[hbp!]
    \centering
    \includegraphics[width=0.46\linewidth]{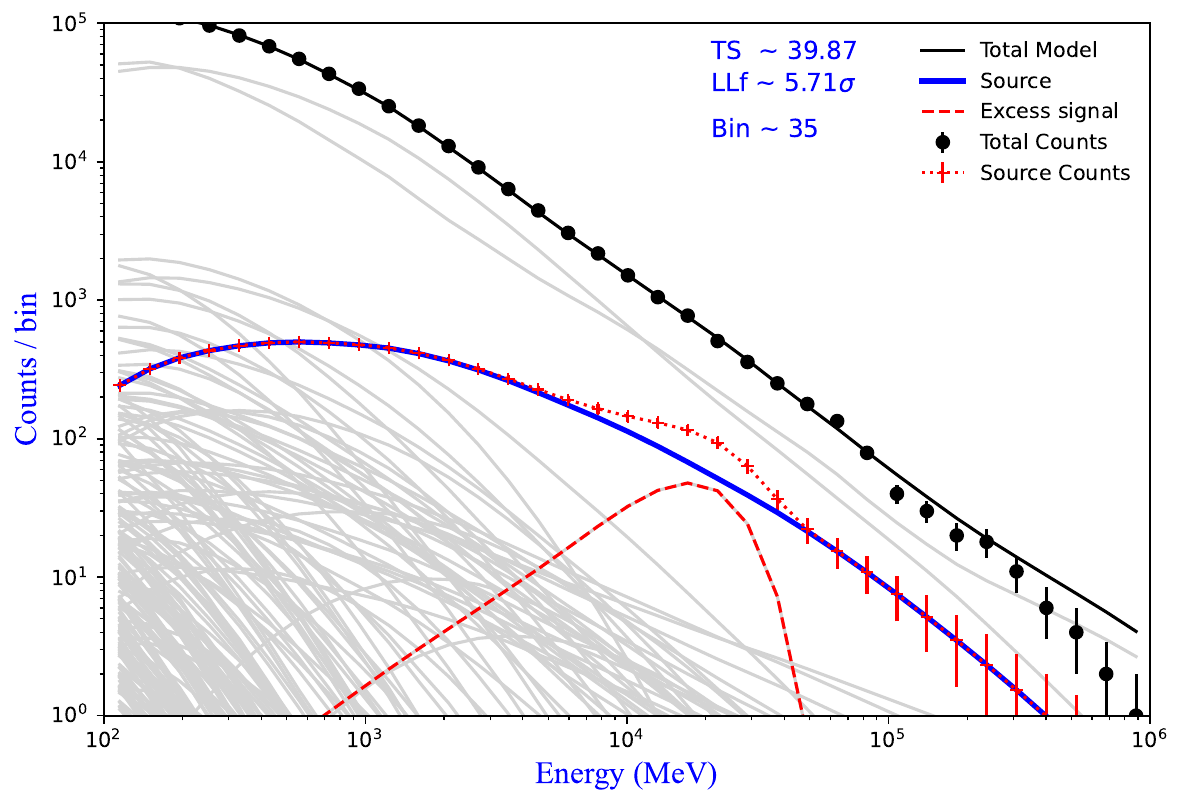}
    \includegraphics[width=0.46\linewidth]{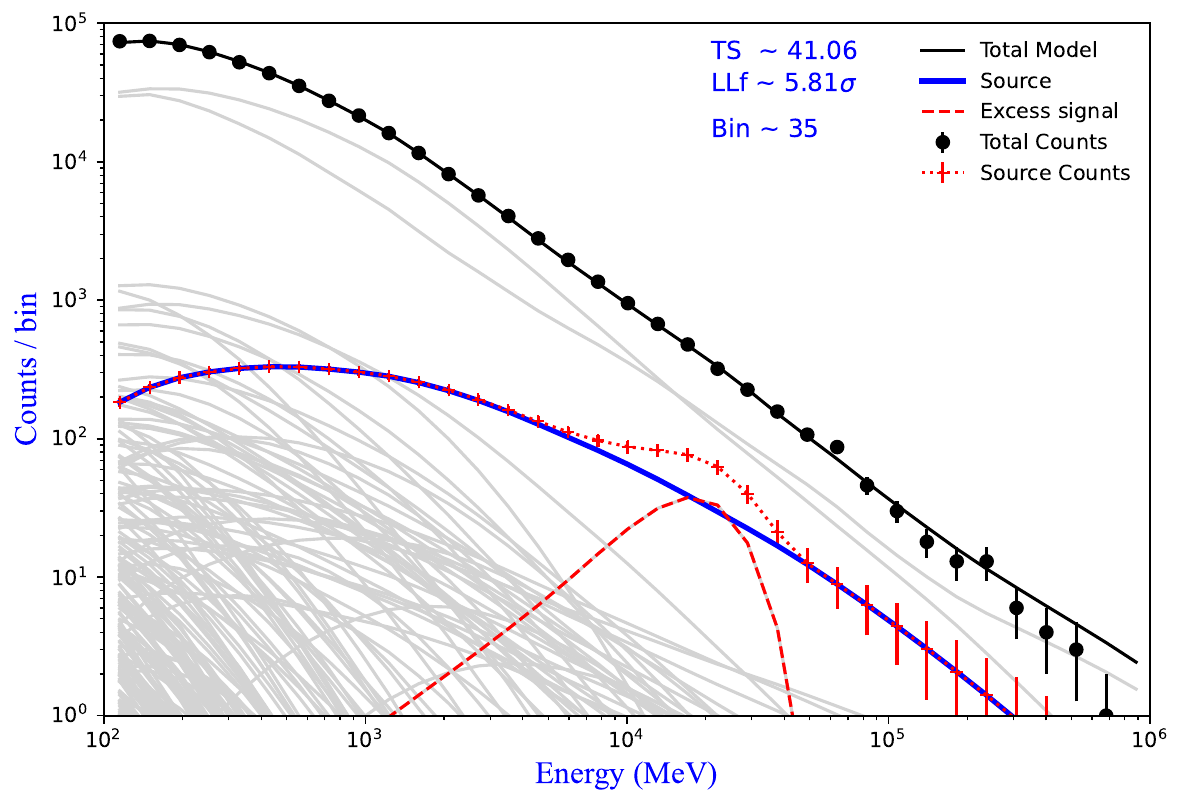}
    \includegraphics[width=0.46\linewidth]{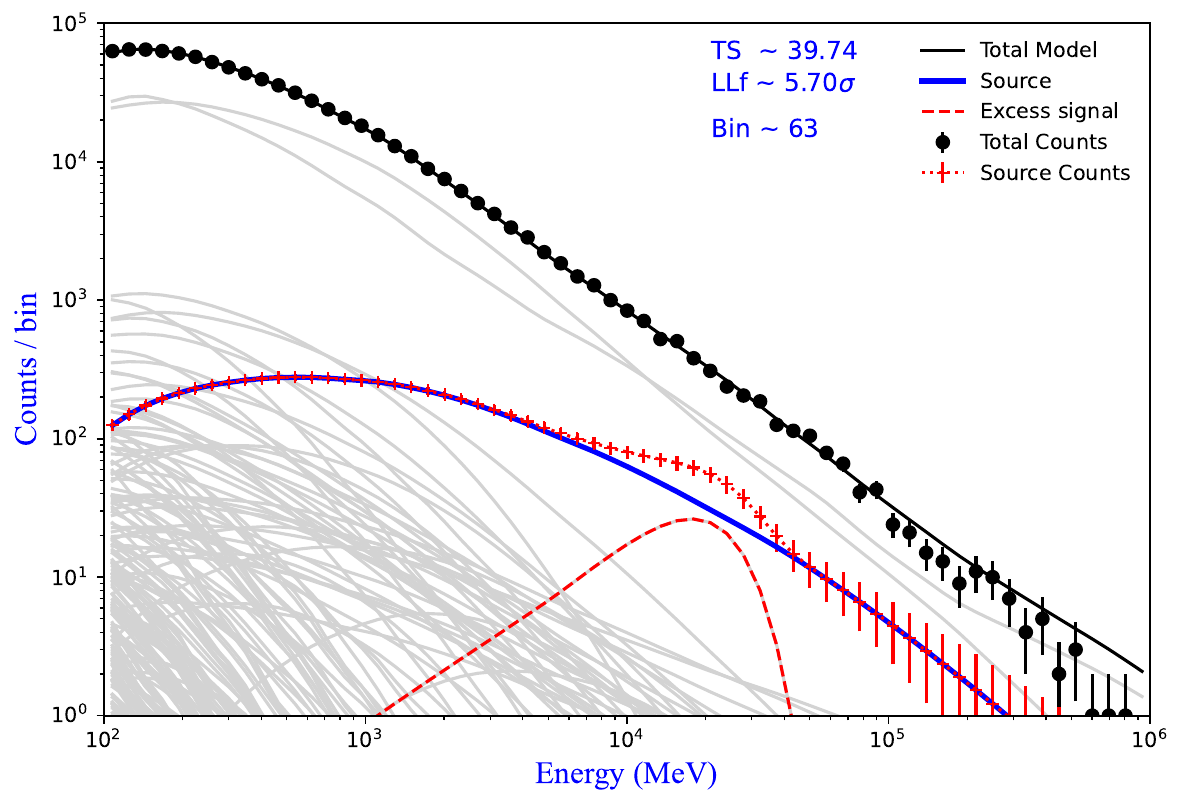}
    \includegraphics[width=0.46\linewidth]{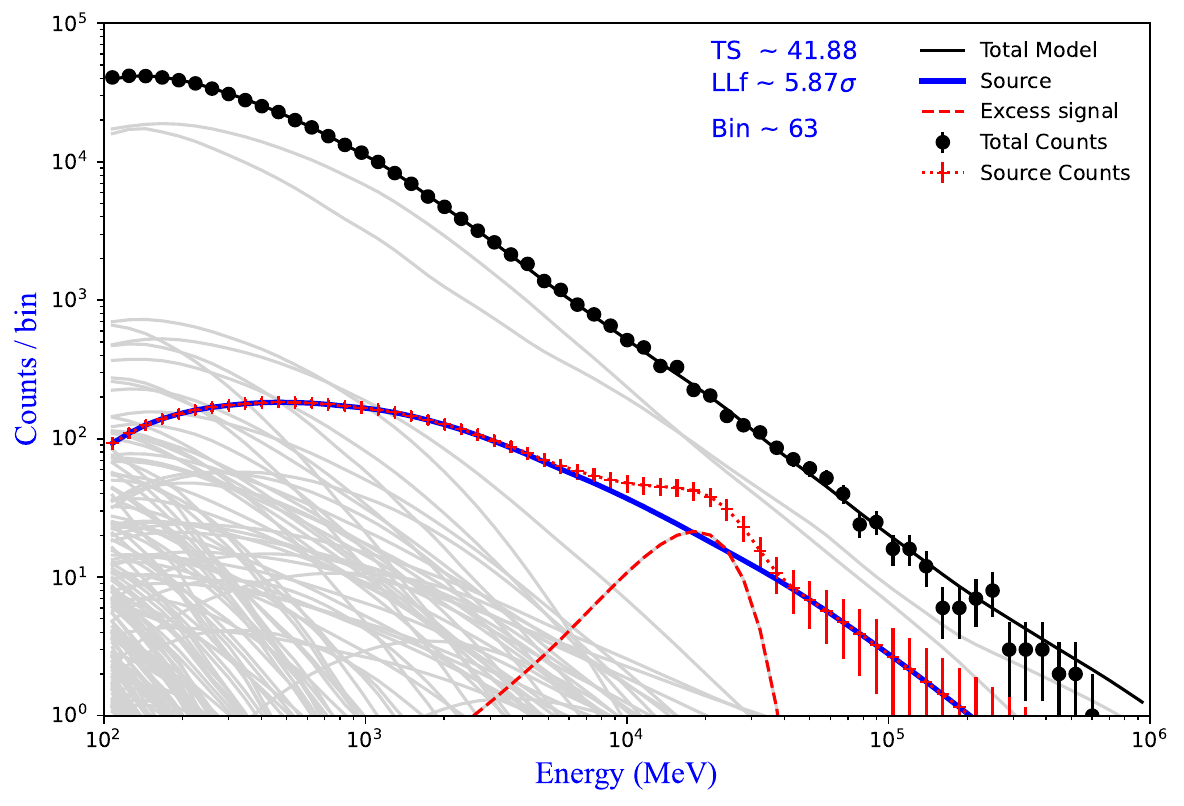}
    \includegraphics[width=0.46\linewidth]{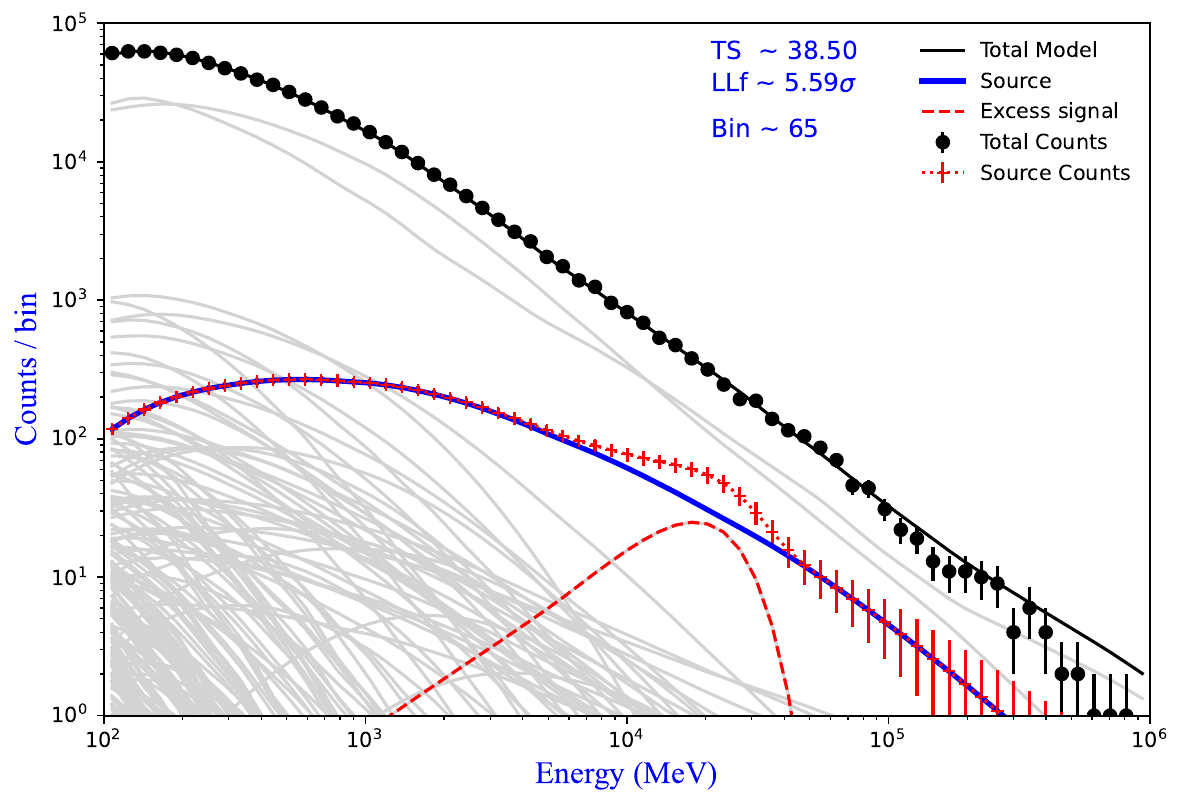}
    \includegraphics[width=0.46\linewidth]{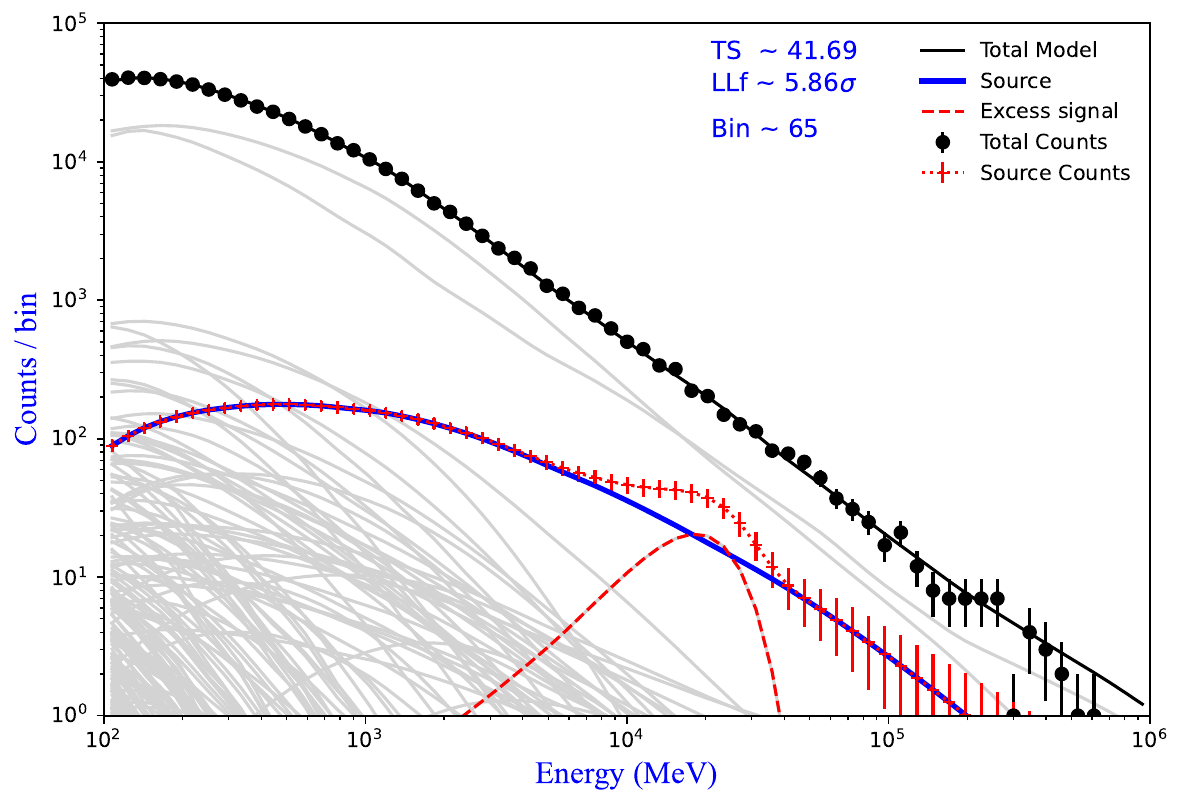}
    \caption{Model-fitted count spectra of 4FGL J0136.5+3906 over {17}-year (left) and 11-year (right) subset with `N' logarithmically spaced energy bins. Black dots and black line: data and total model; light grey lines: background sources; red dot-dashed: source model; blue solid: log-parabola continuum; red dashed: Gaussian excess. The corresponding spectral‑fit parameters are listed in Table \ref{tab1}. Partial spectra are presented herein to provide guidance regarding their structure and content.}
    \label{Fig_count_A1}
\end{figure}

\begin{figure}[hbp!]
    \centering
    \includegraphics[width=0.46\linewidth]{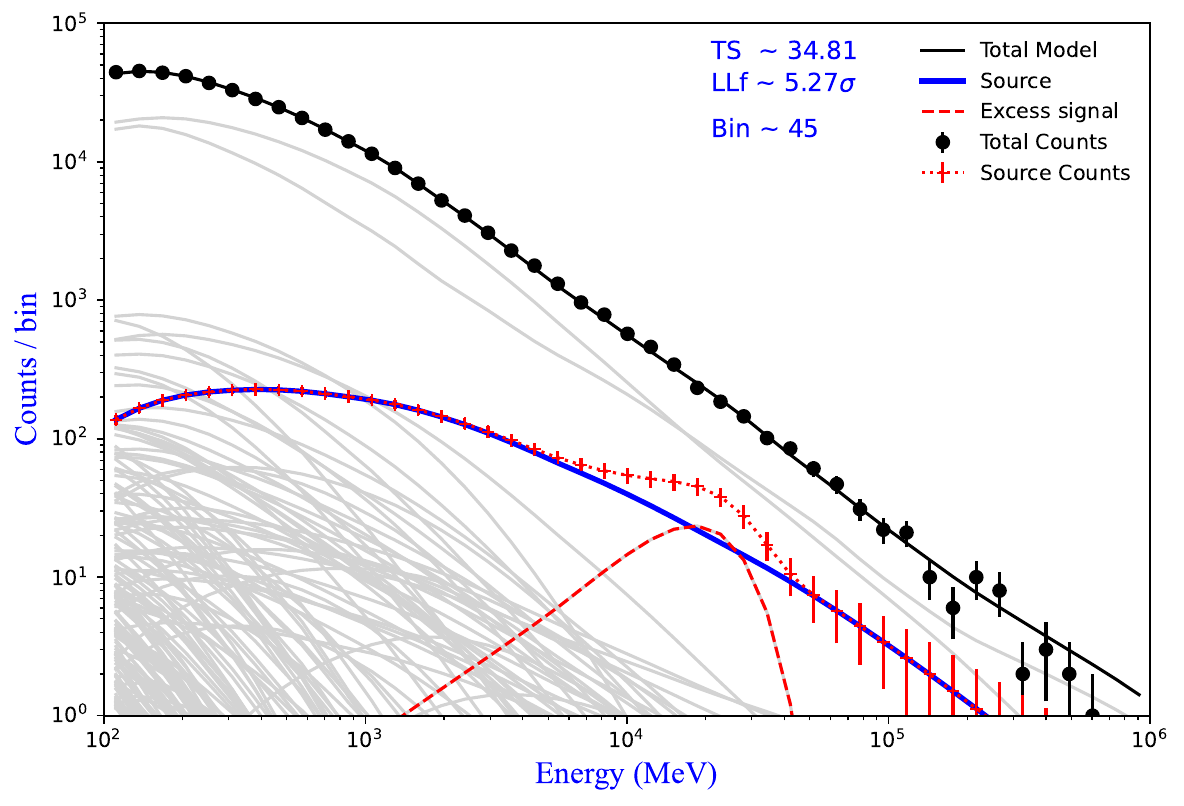}
    \includegraphics[width=0.46\linewidth]{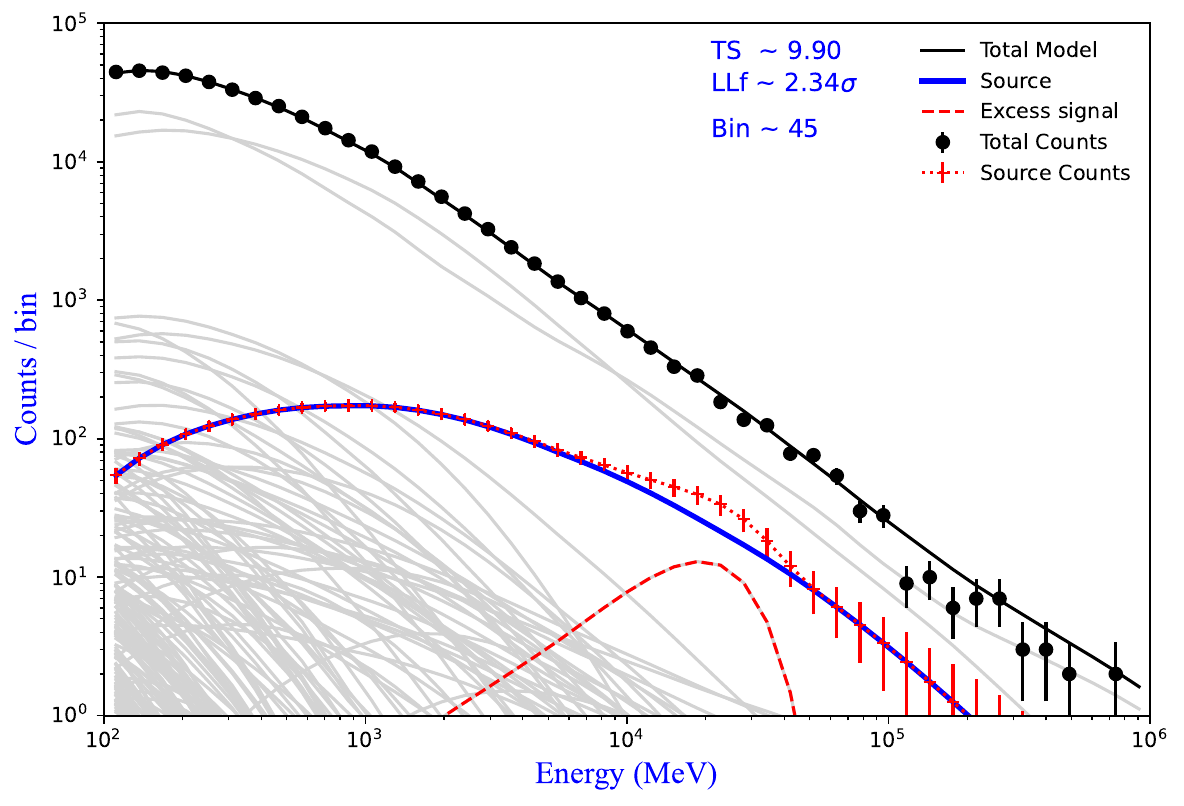}
    \caption{Model-fitted count spectra of 4FGL J0136.5+3906 over a first {8.5}-year (left, 2008.8-2017.2) and a second {8.5}-year (right, 2017.2-2025.8) subset with 45 logarithmically spaced energy bins. Black dots and black line: data and total model; light grey lines: background sources; red dot-dashed: source model; blue solid: log-parabola continuum; red dashed: Gaussian excess. The corresponding spectral‑fit parameters are listed in Table \ref{tab1}.}
    \label{Fig_count_A2}
\end{figure}

\begin{figure*}[htbp!]
\centering
\includegraphics[width=0.96\linewidth]{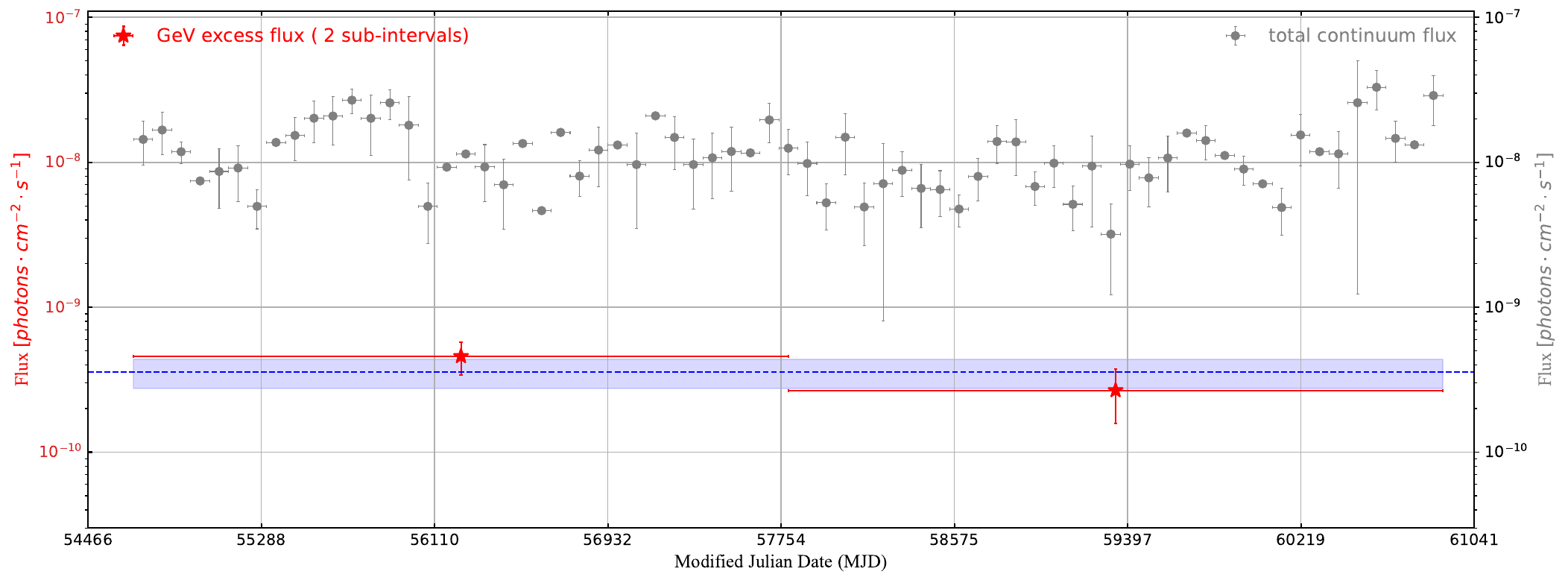}
\caption{\textbf{Temporal stability of the spectral excess for 4FGL J0136.5+3906}. 
The 17-year period (2008.8--2025.8) is split into two equal sub-periods in the 100\,MeV--1\,TeV range. Red stars (left $y$-axis): flux of the spectral excess in each sub-period. Blue dotted line and shaded area: best-fit constant value and its $1\sigma$ uncertainty from fitting a constant to the two red points. Grey (dots) curve (right $y$-axis): total gamma-ray flux above 100\,MeV.}
\label{Fig_Temporal}
\end{figure*}

This appendix presents the complete spectral fitting parameters (Table~\ref{tab1}) and robustness tests (Table~\ref{tab2}), as well as partial model-fitted count spectra for different energy binning schemes and time intervals (Figure~\ref{Fig_count_A1} and \ref{Fig_count_A2}), and temporal stability (Figure~\ref{Fig_Temporal}).

\begin{sidewaystable}
\setlength{\tabcolsep}{1.5pt}
\caption{\textbf{Spectral Fitting Parameters for the {excess} Signal.}}\label{tab1} 
\begin{tabular*}{\textwidth}{@{\extracolsep\fill}lcccccccccc}
\toprule
  &&\multicolumn{4}{c}{Gaussian\footnote{Parameters from the Gaussian signal.}}
  &\multicolumn{5}{c}{Likelihood\footnote{Parameters related to the log-likelihood.}}\\
\cline{3-6}\cline{7-11}
Time interval (MET, s)    &  Bins  &  $E_{\rm peak}$ (GeV) &  $E_{\rm signal}$ (GeV) & $\sigma_{\rm signal}$ (GeV) & $F_{\rm signal}$       & $\mathcal{L}_{\rm null}$  & $\mathcal{L}_{\rm signal}$ & $\mathrm{TS}$  & $\sigma$  & $\sigma_1$\\  
\cline{1-11}
239557417	to 799602221    &  45	&	20.19 $\pm$ 1.66 	& $12.81 \pm 2.62$	 	 	&  $11.03 \pm 0.01$	 	 	&	3.56 $\pm$ 0.73 	&	1209131.29 	&	1209112.47   &	37.63 	&	5.52  & 6.13 \\
239557417	to 794331821    &  45	&	19.92 $\pm$ 2.84 	& $14.03 \pm 4.03$	 	 	&  $10.00 \pm 0.13$	 	 	&	3.36 $\pm$ 1.18 	&	966109.65 	&	966093.88   &	38.40 	&	5.59  & 6.20 \\
\cdashline{1-11}[1pt/1pt]
239557417	to 776015021    &  35	&	18.59 $\pm$ 1.45  	& $12.06 \pm 2.23$	 	 	&  $10.00 \pm 0.04$	 	 	&	3.86 $\pm$ 0.76 	&	991982.82 	&	991962.88   &	39.87 	&	5.71  & 6.31 \\
239557417	to 776015021    &  45*	&	19.26 $\pm$ 1.52 	& $13.07 \pm 2.25$	 	 	&  $10.00 \pm 0.02$	 	 	&	3.62 $\pm$ 0.72 	&	1193520.24	&	1193500.35  &	39.79 	&	5.70  & 6.31 \\
239557417	to 776015021    &  63	&	18.66 $\pm$ 1.45 	& $12.18 \pm 2.22$	 	 	&  $10.00 \pm 0.00$	 	 	&	3.72 $\pm$ 0.74 	&	1463626.99	&	1463607.13  &	39.74  	&	5.70  & 6.30 \\
239557417	to 776015021    &  65	&	19.03 $\pm$ 1.51 	& $12.73 \pm 2.26$	 	 	&  $10.00 \pm 0.00$	 	 	&	3.57 $\pm$ 0.72 	&	1488735.94 	&	1488716.69  &	38.50 	&	5.59  & 6.20 \\
\cdashline{1-11}[1pt/1pt]
239557417	to 681320621    &  45	&	19.53 $\pm$ 3.86 	& $15.88 \pm 3.71$	 	 	&	$8.05 \pm 2.93$	 	 	&	3.53 $\pm$ 1.12 	&	1108274.64 	&	1108254.27  &	40.73 	&	5.78  & 6.38 \\
\cdashline{1-11}[1pt/1pt]
239557417	to	586626221	&	35	&	18.64 $\pm$ 4.07 	&	14.02 	$\pm$	3.99 	&	8.68 	$\pm$	2.95 	&	4.13 $\pm$ 1.39 	&	864884.98 	&	864864.45 	&	41.06 	&	5.81  &	6.41 \\
239557417	to	586626221	&	45**&	19.00 $\pm$ 3.97 	&	14.82 	$\pm$	3.77 	&	8.41 	$\pm$	3.02 	&	3.85 $\pm$ 1.36 	&	994058.74 	&	994038.21 	&	41.07 	&	5.81  &	6.41 \\
239557417	to	586626221	&	48	&	18.56 $\pm$ 3.93 	&	14.29 	$\pm$	3.83 	&	8.37 	$\pm$	2.88 	&	3.92 $\pm$ 1.38 	&	1027276.66 	&	1027256.38 	&	40.55 	&	5.77  &	6.37 \\
239557417	to	586626221	&	51	&	18.90 $\pm$ 3.40 	&	15.13 	$\pm$	3.30 	&	8.01 	$\pm$	2.53 	&	3.61 $\pm$ 1.21 	&	1058506.42 	&	1058486.71 	&	39.42 	&	5.67  &	6.28 \\
239557417	to	586626221	&	53	&	18.15 $\pm$ 3.04 	&	14.56 	$\pm$	2.95 	&	7.66 	$\pm$	2.25 	&	3.72 $\pm$ 1.18 	&  	1078279.51 	&	1078259.17 	&	40.69 	&	5.78  &	6.38 \\
239557417	to	586626221	&	56	&	18.67 $\pm$ 3.74 	&	14.30 	$\pm$	3.72 	&	8.49 	$\pm$	2.67 	&	3.83 $\pm$ 1.29 	&	1106625.43 	&	1106605.85 	&	39.17 	&	5.65  &	6.26 \\
239557417	to	586626221	&	62	&	19.05 $\pm$ 4.02 	&	14.43 	$\pm$	3.96 	&	8.79 	$\pm$	2.90 	&	3.87 $\pm$ 1.32 	&	1159001.58 	&	1158981.60 	&	39.98 	&	5.72  &	6.32 \\
239557417	to	586626221	&	63	&	18.60 $\pm$ 3.64  	&	14.95 	$\pm$	3.42 	&	7.83 	$\pm$	2.84 	&	3.69 $\pm$ 1.35 	&	1167274.81 	&	1167253.87 	&	41.88 	&	5.87  &	6.47 \\
239557417	to	586626221	&	65	&	18.79 $\pm$ 3.61 	&	14.72 	$\pm$	3.50 	&	8.26 	$\pm$	2.67 	&	3.81 $\pm$ 1.29 	&	1183355.10 	&	1183334.26 	&	41.69 	&	5.86  &	6.46 \\
\cdashline{1-11}[1pt/1pt]
239557417	to 578677421    &  45	&	18.64 $\pm$ 3.38 	& $14.91 \pm 3.32$	 	 	&	$7.91 \pm 2.53$	 	 	&	3.71 $\pm$ 1.32 	&	981737.63 	&	981717.14  &	40.98 	&	5.80  & 6.40 \\
\cdashline{1-11}[1pt/1pt]
239557417	to 507915821    &  45	&	18.92 $\pm$ 2.37 	& $13.08 \pm 2.93$	 	 	&	$9.67 \pm 1.19$	 	  	&	4.59 $\pm$ 1.17 	&	871432.62 	&	871415.21  &	34.81 	&	5.27  & 5.90 \\
507915821	to 776015021    &  45	&	20.36 $\pm$ 3.39 	& $13.08 \pm 5.28$	 	 	&	$11.03 \pm 0.01$	 	&	2.67 $\pm$ 1.09 	&	878983.70 	&	878978.75  &	9.90  	&	2.34  & 3.15 \\
\cdashline{1-11}[1pt/1pt]
239557417	to 373736619    &  45	&	18.63 $\pm$ 3.08  	& $10.17 \pm 5.64$	 	 	&	$11.03 \pm 0.01$	 	&	3.93 $\pm$ 1.68 	&	597167.43 	&	597163.12  &	8.63 	&	2.11  & 2.93  \\
373736619	to 507915821    &  45	&	19.31 $\pm$ 3.78   	& $15.54 \pm 3.90$	 	 	&	$8.09 \pm 2.51$	    	&	4.98 $\pm$ 1.80 	&	550019.17 	&	550003.06   &	32.23 	&	5.04  & 5.67  \\
507915821	to 641835821    &  45	&	18.65 $\pm$ 2.67   	& $18.01 \pm 2.55$	 	 	&	$3.40 \pm 2.78$ 	 	&	1.40 $\pm$ 0.86 	&	567181.57 	&	567177.82  &	7.51 	&	1.90  & 2.74 \\
641835821	to 776015021    &  45	&	22.06 $\pm$ 49.95   & $16.93 \pm 6.51$	 	 	&	$10.00 \pm 0.01$	    &	2.06 $\pm$ 8.89 	&	592023.52 	&	592021.42   &	4.20 	&	1.17  & 2.05  \\
\bottomrule
\end{tabular*}
\par\noindent Note: 
Column 1: time interval (e.g., 
MET 239557417-799602221, covering $\sim$17.75 years from 2008 August 4 to 2026 May 4; 
MET 239557417-794331821, covering $\sim$17.6 years from 2008 August 4 to 2026 March 4; 
MET 239557417-776015021, 17 years, covering $\sim$17 years from 2008 August 4 to 2025 August 4;
MET 239557417-681320621, 14 years;
MET 239557417-586626221, 11 years; 
MET 239557417-507915821, 8.5 years, from 2008.8 to 2017.2;
MET 507915821-776015021, 8.5 years, from 2017.2 to 2025.8,
respectively). 
Column 2: The entire energy spectrum ranging from 100 MeV to 1 TeV is systematically divided into `N bins' equidistant energy intervals on a logarithmic scale. 
Columns 3: $E_{\rm peak}$ is the peak of the excess in logarithmic energy scale.
Columns 4 \& 5: Center energy and width of the fitting parameters for the Gaussian excess in linear energy space. 
Columns 6: The flux of excess signal ($\times{\rm 10^{-10}~ph~cm^{-2}~s^{-1}}$).
Columns 7 \& 8: Log-likelihood values for the null and signal models. 
Columns 9 \& 10: Test statistic and corresponding local significance ($\sigma$) for three degrees of freedom. 
Column 11: Significance for one degree of freedom ($\sigma_1$).
Here, the Model-fitted count spectra for 45 bins (marked asterisk*,**) is 
presented in Figure \ref{fig_count} to provide guidance regarding its form and content.
\end{sidewaystable}

\clearpage

\begin{table*}[hbtp!]
\caption{Robustness Tests for the Spectral excess}\label{tab2}
\begin{tabular*}{\textwidth}{lcccccc}
\toprule
Analysis Case                 &  $\sigma$ & TS  & $E_{\rm peak}$ (GeV) & $F_{\rm signal} (\times 10^{-10})$  & Note                    \\
\midrule
\multirow{2}{*}{\centering Baseline Analysis (45** bins) }   & 
\multirow{2}{*}{\centering 5.81 }               & 
\multirow{2}{*}{\centering 41.07 }              &  
\multirow{2}{*}{\centering 19.00 $\pm$ 3.97 }   & 
\multirow{2}{*}{\centering 3.85 $\pm$ 1.36  }   &                                          Uses P8R3\_SOURCE\_V3 IRF                     \\
                              &       &         &                   &                    & and gll\_iem\_v07 background model.           \\   
\cdashline{1-6}[1pt/1pt]
{IRF Systematics}               &        &        &                   &                    &                               \\
\cdashline{1-5}[1pt/1pt]
--- P8R3\_SOURCE\_V2          & 5.81   & 41.07  & 19.00 $\pm$ 3.97  & 3.85 $\pm$ 1.36 &          peak of the excess signal                \\
--- P8R3\_CLEAN\_V3           & 6.05   & 44.06  & 18.68 $\pm$ 3.94  & 4.38 $\pm$ 1.47 &          stable at $\sim$ 19 GeV.                 \\
--- P8R3\_SOURCEVETO\_V3      & 5.78   & 40.77  & 19.08 $\pm$ 3.69  & 3.97 $\pm$ 1.37 &                                                   \\
\cdashline{1-5}[1pt/1pt]
{edisp Systematics}             &        &        &                   &                 &                                  \\  
\cdashline{1-5}[1pt/1pt]
--- edisp\_bins=-2            & 5.81   & 41.04  & 19.34 $\pm$ 4.22  & 3.89 $\pm$ 1.43 &                                                   \\
\cdashline{1-5}[1pt/1pt]
{Aeff Systematics}             &        &        &                   &                 &                                   \\  
\cdashline{1-5}[1pt/1pt]
--- Bracketing Aeff (min)            & 5.80   & 40.92  & 19.09 $\pm$ 3.67   & 3.71 $\pm$ 1.28 &          \\
--- Bracketing Aeff (max)            & 5.80   & 40.97  & 18.97 $\pm$ 3.81   & 3.84 $\pm$ 1.32 &          \\
--- Bracketing Aeff (min,noedisp)    & 5.80   & 40.91  & 19.12 $\pm$ 3.60   & 3.68 $\pm$ 1.26 &          \\
--- Bracketing Aeff (max,noedisp)    & 5.80   & 40.98  & 18.95 $\pm$ 3.84   & 3.86 $\pm$ 1.33 &          \\
\cdashline{1-5}[1pt/1pt]
{Background Model Systematics}  &        &        &                   &                 &                                                \\
\cdashline{1-5}[1pt/1pt]
--- Diffuse Norm $\times$ 1.1 & 5.81   & 41.07  & 19.00 $\pm$ 3.97  & 3.85 $\pm$ 1.36 &       excess flux stable at                                     \\
--- Diffuse Norm $\times$ 0.9 & 5.81   & 41.07  & 19.00 $\pm$ 3.97  & 3.85 $\pm$ 1.36 &       $\sim$[3.5,4.4] $\times 10^{-10}$ ph cm$^{-2}$ s$^{-1}$   \\
\cdashline{1-5}[1pt/1pt]
{Summed Likelihood joint Analysis}    &        &        &                   &                 &                                           \\
\cdashline{1-5}[1pt/1pt]
--- fount+back                & 5.78   & 40.73  & 19.15 $\pm$ 3.20  & 3.57 $\pm$ 1.17  &      significance stable at                                      \\
--- PSF0$-$PSF3               & 5.96   & 43.01  & 19.16 $\pm$ 3.21  & 3.72 $\pm$ 1.18  &      $\sim$[5.7,6.1]$\sigma$                                     \\
--- {EDISP0$-$EDISP3}          & 5.81   & 41.06  & 19.09 $\pm$ 3.29  & 3.70 $\pm$ 1.21  &       TS $\sim$ [40,44]                          \\
\bottomrule
\end{tabular*}
\par\noindent Note: The Baseline Analysis based one the 45** bins (see Table \ref{tab1}) for MET 239557417-586626221, 11 year of period  from 2008 August 4 to 2019 August 4.
The normalization parameters of the Galactic and isotropic diffuse components were left free in all fits (see Section~2.2). Changing the input diffuse model normalization by $\pm10\%$ therefore results in the same best-fit values, because the likelihood re-optimizes the diffuse normalizations. This confirms that the spectral excess is robust against diffuse modeling uncertainties and other systematic uncertainties. For the IRF, edisp and summed likelihood tests, the parameters also remain stable. Under P8R3\_SOURCE\_V2, Diffuse Norm $\times$ 1.1, and Diffuse Norm $\times$ 0.9, the fitting parameters converge to the same values as for P8R3\_SOURCE\_V3. See LAT Data Products \url{https://fermi.gsfc.nasa.gov/ssc/data/access/lat/lat_data_products.html} and LAT Background Models \url{https://fermi.gsfc.nasa.gov/ssc/data/access/lat/BackgroundModels.html} and the Bracketing Aeff \url{https://fermi.gsfc.nasa.gov/ssc/data/analysis/scitools/Aeff_Systematics.html}.
\end{table*}

\bibliography{sample701}{}

@ARTICLE{2026arXiv260928303K,
       author = {{Kang}, Shi-Ju and {Yin}, Yue and {Zheng}, Yong-Gang and {Wu}, Qingwen},
        title = "{A Tentative Line-like GeV Excess in Fermi Blazar 4FGL J1754.2+3212: Implications for Jet Physics and Beyond}",
      journal = {arXiv e-prints},
         year = 2026,
        month = sep,
          eid = {arXiv:2609.28303},
        pages = {arXiv:2609.28303},
archivePrefix = {arXiv},
       eprint = {2609.28303},
 primaryClass = {astro-ph.HE},
       adsurl = {https://ui.adsabs.harvard.edu/abs/2026arXiv260928303K}
}

@ARTICLE{2026arXiv260623488K,
       author = {{Kang}, Shi-Ju},
        title = "{A Tentative Double Excess in the Gamma-Ray Spectrum of the Fermi Blazar 4FGL J0604.9-0000}",
      journal = {arXiv e-prints},
         year = 2026,
        month = jun,
          eid = {arXiv:2606.23488},
        pages = {arXiv:2606.23488},
          doi = {10.48550/arXiv.2606.23488},
archivePrefix = {arXiv},
       eprint = {2606.23488},
 primaryClass = {astro-ph.HE},
       adsurl = {https://ui.adsabs.harvard.edu/abs/2026arXiv260623488K}
}

@ARTICLE{2012ApJS..203....4A,
       author = {{Ackermann}, M. and {Ajello}, M. and {Albert}, A. and {Allafort}, A. and {Atwood}, W.~B. and {Axelsson}, M. and {Baldini}, L. and {Ballet}, J. and {Barbiellini}, G. and {Bastieri}, D. and {Bechtol}, K. and {Bellazzini}, R. and {Bissaldi}, E. and {Blandford}, R.~D. and {Bloom}, E.~D. and {Bogart}, J.~R. and {Bonamente}, E. and {Borgland}, A.~W. and {Bottacini}, E. and {Bouvier}, A. and {Brandt}, T.~J. and {Bregeon}, J. and {Brigida}, M. and {Bruel}, P. and {Buehler}, R. and {Burnett}, T.~H. and {Buson}, S. and {Caliandro}, G.~A. and {Cameron}, R.~A. and {Caraveo}, P.~A. and {Casandjian}, J.~M. and {Cavazzuti}, E. and {Cecchi}, C. and {{\c{C}}elik}, {\"O}. and {Charles}, E. and {Chaves}, R.~C.~G. and {Chekhtman}, A. and {Cheung}, C.~C. and {Chiang}, J. and {Ciprini}, S. and {Claus}, R. and {Cohen-Tanugi}, J. and {Conrad}, J. and {Corbet}, R. and {Cutini}, S. and {D'Ammando}, F. and {Davis}, D.~S. and {de Angelis}, A. and {DeKlotz}, M. and {de Palma}, F. and {Dermer}, C.~D. and {Digel}, S.~W. and {Silva}, E. do Couto e. and {Drell}, P.~S. and {Drlica-Wagner}, A. and {Dubois}, R. and {Favuzzi}, C. and {Fegan}, S.~J. and {Ferrara}, E.~C. and {Focke}, W.~B. and {Fortin}, P. and {Fukazawa}, Y. and {Funk}, S. and {Fusco}, P. and {Gargano}, F. and {Gasparrini}, D. and {Gehrels}, N. and {Giebels}, B. and {Giglietto}, N. and {Giordano}, F. and {Giroletti}, M. and {Glanzman}, T. and {Godfrey}, G. and {Grenier}, I.~A. and {Grove}, J.~E. and {Guiriec}, S. and {Hadasch}, D. and {Hayashida}, M. and {Hays}, E. and {Horan}, D. and {Hou}, X. and {Hughes}, R.~E. and {Jackson}, M.~S. and {Jogler}, T. and {J{\'o}hannesson}, G. and {Johnson}, R.~P. and {Johnson}, T.~J. and {Johnson}, W.~N. and {Kamae}, T. and {Katagiri}, H. and {Kataoka}, J. and {Kerr}, M. and {Kn{\"o}dlseder}, J. and {Kuss}, M. and {Lande}, J. and {Larsson}, S. and {Latronico}, L. and {Lavalley}, C. and {Lemoine-Goumard}, M. and {Longo}, F. and {Loparco}, F. and {Lott}, B. and {Lovellette}, M.~N. and {Lubrano}, P. and {Mazziotta}, M.~N. and {McConville}, W. and {McEnery}, J.~E. and {Mehault}, J. and {Michelson}, P.~F. and {Mitthumsiri}, W. and {Mizuno}, T. and {Moiseev}, A.~A. and {Monte}, C. and {Monzani}, M.~E. and {Morselli}, A. and {Moskalenko}, I.~V. and {Murgia}, S. and {Naumann-Godo}, M. and {Nemmen}, R. and {Nishino}, S. and {Norris}, J.~P. and {Nuss}, E. and {Ohno}, M. and {Ohsugi}, T. and {Okumura}, A. and {Omodei}, N. and {Orienti}, M. and {Orlando}, E. and {Ormes}, J.~F. and {Paneque}, D. and {Panetta}, J.~H. and {Perkins}, J.~S. and {Pesce-Rollins}, M. and {Pierbattista}, M. and {Piron}, F. and {Pivato}, G. and {Porter}, T.~A. and {Racusin}, J.~L. and {Rain{\`o}}, S. and {Rando}, R. and {Razzano}, M. and {Razzaque}, S. and {Reimer}, A. and {Reimer}, O. and {Reposeur}, T. and {Reyes}, L.~C. and {Ritz}, S. and {Rochester}, L.~S. and {Romoli}, C. and {Roth}, M. and {Sadrozinski}, H.~F.-W. and {Sanchez}, D.~A. and {Saz Parkinson}, P.~M. and {Sbarra}, C. and {Scargle}, J.~D. and {Sgr{\`o}}, C. and {Siegal-Gaskins}, J. and {Siskind}, E.~J. and {Spandre}, G. and {Spinelli}, P. and {Stephens}, T.~E. and {Suson}, D.~J. and {Tajima}, H. and {Takahashi}, H. and {Tanaka}, T. and {Thayer}, J.~G. and {Thayer}, J.~B. and {Thompson}, D.~J. and {Tibaldo}, L. and {Tinivella}, M. and {Tosti}, G. and {Troja}, E. and {Usher}, T.~L. and {Vandenbroucke}, J. and {Van Klaveren}, B. and {Vasileiou}, V. and {Vianello}, G. and {Vitale}, V. and {Waite}, A.~P. and {Wallace}, E. and {Winer}, B.~L. and {Wood}, D.~L. and {Wood}, K.~S. and {Wood}, M. and {Yang}, Z. and {Zimmer}, S.},
        title = "{The Fermi Large Area Telescope on Orbit: Event Classification, Instrument Response Functions, and Calibration}",
      journal = {\apjs},
         year = 2012,
        month = nov,
       volume = {203},
       number = {1},
          eid = {4},
        pages = {4},
          doi = {10.1088/0067-0049/203/1/4},
archivePrefix = {arXiv},
       eprint = {1206.1896},
 primaryClass = {astro-ph.IM},
       adsurl = {https://ui.adsabs.harvard.edu/abs/2012ApJS..203....4A}
}

@ARTICLE{2014JCAP...10..023A,
       author = {{Albert}, Andrea and {G{\'o}mez-Vargas}, Germ{\'a}n A. and {Grefe}, Michael and {Mu{\~n}oz}, Carlos and {Weniger}, Christoph and {Bloom}, Elliott D. and {Charles}, Eric and {Mazziotta}, Mario N. and {Morselli}, Aldo},
        title = "{Search for 100 MeV to 10 GeV {\ensuremath{\gamma}}-ray lines in the Fermi-LAT data and implications for gravitino dark matter in the {\ensuremath{\mu}}{\ensuremath{\nu}}SSM}",
      journal = {\jcap},
         year = 2014,
        month = oct,
       volume = {2014},
       number = {10},
        pages = {023-023},
          doi = {10.1088/1475-7516/2014/10/023},
archivePrefix = {arXiv},
       eprint = {1406.3430},
 primaryClass = {astro-ph.HE},
       adsurl = {https://ui.adsabs.harvard.edu/abs/2014JCAP...10..023A}
}

@ARTICLE{2013PhRvD..88h2002A,
       author = {{Ackermann}, M. and {Ajello}, M. and {Albert}, A. and {Allafort}, A. and {Baldini}, L. and {Barbiellini}, G. and {Bastieri}, D. and {Bechtol}, K. and {Bellazzini}, R. and {Bissaldi}, E. and {Bloom}, E.~D. and {Bonamente}, E. and {Bottacini}, E. and {Brandt}, T.~J. and {Bregeon}, J. and {Brigida}, M. and {Bruel}, P. and {Buehler}, R. and {Buson}, S. and {Caliandro}, G.~A. and {Cameron}, R.~A. and {Caraveo}, P.~A. and {Casandjian}, J.~M. and {Cecchi}, C. and {Charles}, E. and {Chaves}, R.~C.~G. and {Chekhtman}, A. and {Chiang}, J. and {Ciprini}, S. and {Claus}, R. and {Cohen-Tanugi}, J. and {Conrad}, J. and {Cutini}, S. and {D'Ammando}, F. and {de Angelis}, A. and {de Palma}, F. and {Dermer}, C.~D. and {Digel}, S.~W. and {Di Venere}, L. and {Drell}, P.~S. and {Drlica-Wagner}, A. and {Essig}, R. and {Favuzzi}, C. and {Fegan}, S.~J. and {Ferrara}, E.~C. and {Focke}, W.~B. and {Franckowiak}, A. and {Fukazawa}, Y. and {Funk}, S. and {Fusco}, P. and {Gargano}, F. and {Gasparrini}, D. and {Germani}, S. and {Giglietto}, N. and {Giordano}, F. and {Giroletti}, M. and {Glanzman}, T. and {Godfrey}, G. and {Gomez-Vargas}, G.~A. and {Grenier}, I.~A. and {Guiriec}, S. and {Gustafsson}, M. and {Hadasch}, D. and {Hayashida}, M. and {Hill}, A.~B. and {Horan}, D. and {Hou}, X. and {Hughes}, R.~E. and {Inoue}, Y. and {Izaguirre}, E. and {Jogler}, T. and {Kamae}, T. and {Kn{\"o}dlseder}, J. and {Kuss}, M. and {Lande}, J. and {Larsson}, S. and {Latronico}, L. and {Longo}, F. and {Loparco}, F. and {Lovellette}, M.~N. and {Lubrano}, P. and {Malyshev}, D. and {Mayer}, M. and {Mazziotta}, M.~N. and {McEnery}, J.~E. and {Michelson}, P.~F. and {Mitthumsiri}, W. and {Mizuno}, T. and {Moiseev}, A.~A. and {Monzani}, M.~E. and {Morselli}, A. and {Moskalenko}, I.~V. and {Murgia}, S. and {Nakamori}, T. and {Nemmen}, R. and {Nuss}, E. and {Ohsugi}, T. and {Okumura}, A. and {Omodei}, N. and {Orienti}, M. and {Orlando}, E. and {Ormes}, J.~F. and {Paneque}, D. and {Perkins}, J.~S. and {Pesce-Rollins}, M. and {Piron}, F. and {Pivato}, G. and {Rain{\`o}}, S. and {Rando}, R. and {Razzano}, M. and {Razzaque}, S. and {Reimer}, A. and {Reimer}, O. and {Romani}, R.~W. and {S{\'a}nchez-Conde}, M. and {Schulz}, A. and {Sgr{\`o}}, C. and {Siegal-Gaskins}, J. and {Siskind}, E.~J. and {Snyder}, A. and {Spandre}, G. and {Spinelli}, P. and {Suson}, D.~J. and {Tajima}, H. and {Takahashi}, H. and {Thayer}, J.~G. and {Thayer}, J.~B. and {Tibaldo}, L. and {Tinivella}, M. and {Tosti}, G. and {Troja}, E. and {Uchiyama}, Y. and {Usher}, T.~L. and {Vandenbroucke}, J. and {Vasileiou}, V. and {Vianello}, G. and {Vitale}, V. and {Winer}, B.~L. and {Wood}, K.~S. and {Wood}, M. and {Yang}, Z. and {Zaharijas}, G. and {Zimmer}, S.},
        title = "{Search for gamma-ray spectral lines with the Fermi Large Area Telescope and dark matter implications}",
      journal = {\prd},
         year = 2013,
        month = oct,
       volume = {88},
       number = {8},
          eid = {082002},
        pages = {082002},
          doi = {10.1103/PhysRevD.88.082002},
archivePrefix = {arXiv},
       eprint = {1305.5597},
 primaryClass = {astro-ph.HE},
       adsurl = {https://ui.adsabs.harvard.edu/abs/2013PhRvD..88h2002A}
}

@ARTICLE{2015PhRvD..91l2002A,
       author = {{Ackermann}, M. and {Ajello}, M. and {Albert}, A. and {Anderson}, B. and {Atwood}, W.~B. and {Baldini}, L. and {Barbiellini}, G. and {Bastieri}, D. and {Bellazzini}, R. and {Bissaldi}, E. and {Blandford}, R.~D. and {Bloom}, E.~D. and {Bonino}, R. and {Bottacini}, E. and {Brandt}, T.~J. and {Bregeon}, J. and {Bruel}, P. and {Buehler}, R. and {Buson}, S. and {Caliandro}, G.~A. and {Cameron}, R.~A. and {Caputo}, R. and {Caragiulo}, M. and {Caraveo}, P.~A. and {Cecchi}, C. and {Charles}, E. and {Chekhtman}, A. and {Chiang}, J. and {Chiaro}, G. and {Ciprini}, S. and {Claus}, R. and {Cohen-Tanugi}, J. and {Conrad}, J. and {Cuoco}, A. and {Cutini}, S. and {D'Ammando}, F. and {de Angelis}, A. and {de Palma}, F. and {Desiante}, R. and {Digel}, S.~W. and {Di Venere}, L. and {Drell}, P.~S. and {Drlica-Wagner}, A. and {Favuzzi}, C. and {Fegan}, S.~J. and {Franckowiak}, A. and {Fukazawa}, Y. and {Funk}, S. and {Fusco}, P. and {Gargano}, F. and {Gasparrini}, D. and {Giglietto}, N. and {Giordano}, F. and {Giroletti}, M. and {Godfrey}, G. and {Gomez-Vargas}, G.~A. and {Grenier}, I.~A. and {Grove}, J.~E. and {Guiriec}, S. and {Gustafsson}, M. and {Hewitt}, J.~W. and {Hill}, A.~B. and {Horan}, D. and {J{\'o}hannesson}, G. and {Johnson}, R.~P. and {Kuss}, M. and {Larsson}, S. and {Latronico}, L. and {Li}, J. and {Li}, L. and {Longo}, F. and {Loparco}, F. and {Lovellette}, M.~N. and {Lubrano}, P. and {Malyshev}, D. and {Mayer}, M. and {Mazziotta}, M.~N. and {McEnery}, J.~E. and {Michelson}, P.~F. and {Mizuno}, T. and {Moiseev}, A.~A. and {Monzani}, M.~E. and {Morselli}, A. and {Murgia}, S. and {Nuss}, E. and {Ohsugi}, T. and {Orienti}, M. and {Orlando}, E. and {Ormes}, J.~F. and {Paneque}, D. and {Pesce-Rollins}, M. and {Piron}, F. and {Pivato}, G. and {Rain{\`o}}, S. and {Rando}, R. and {Razzano}, M. and {Reimer}, A. and {Reposeur}, T. and {Ritz}, S. and {S{\'a}nchez-Conde}, M. and {Schulz}, A. and {Sgr{\`o}}, C. and {Siskind}, E.~J. and {Spada}, F. and {Spandre}, G. and {Spinelli}, P. and {Tajima}, H. and {Takahashi}, H. and {Thayer}, J.~B. and {Tibaldo}, L. and {Torres}, D.~F. and {Tosti}, G. and {Troja}, E. and {Vianello}, G. and {Werner}, M. and {Winer}, B.~L. and {Wood}, K.~S. and {Wood}, M. and {Zaharijas}, G. and {Zimmer}, S. and {Fermi LAT Collaboration}},
        title = "{Updated search for spectral lines from Galactic dark matter interactions with pass 8 data from the Fermi Large Area Telescope}",
      journal = {\prd},
         year = 2015,
        month = jun,
       volume = {91},
       number = {12},
          eid = {122002},
        pages = {122002},
          doi = {10.1103/PhysRevD.91.122002},
archivePrefix = {arXiv},
       eprint = {1506.00013},
 primaryClass = {astro-ph.HE},
       adsurl = {https://ui.adsabs.harvard.edu/abs/2015PhRvD..91l2002A}
}

@ARTICLE{2026ApJ..1002..199Z,
       author = {{Zhao}, X.~Z. and {Pu}, Y.~M. and {Zheng}, Y.~G. and {Kang}, S.~J.},
        title = "{Jet Power and Electron Cooling in Fermi Blazars: Constraints from One-zone Leptonic Spectral Energy Distribution Modeling}",
      journal = {\apj},
         year = 2026,
        month = may,
       volume = {1002},
       number = {2},
          eid = {199},
        pages = {199},
          doi = {10.3847/1538-4357/ae5e5e},
       adsurl = {https://ui.adsabs.harvard.edu/abs/2026ApJ..1002..199Z}
}

@ARTICLE{2014ApJ...783...83L,
       author = {{Liao}, N.~H. and {Bai}, J.~M. and {Liu}, H.~T. and {Weng}, S.~S. and {Chen}, Liang and {Li}, F.},
        title = "{Multiwavelength Variability Properties of Fermi Blazar S5 0716+714}",
      journal = {\apj},
         year = 2014,
        month = mar,
       volume = {783},
       number = {2},
          eid = {83},
        pages = {83},
          doi = {10.1088/0004-637X/783/2/83},
archivePrefix = {arXiv},
       eprint = {1401.4937},
 primaryClass = {astro-ph.HE},
       adsurl = {https://ui.adsabs.harvard.edu/abs/2014ApJ...783...83L}
}

@ARTICLE{2008ApJ...684L..73A,
       author = {{Acciari}, V.~A. and {Aliu}, E. and {Beilicke}, M. and {Benbow}, W. and {B{\"o}ttcher}, M. and {Bradbury}, S.~M. and {Buckley}, J.~H. and {Bugaev}, V. and {Butt}, Y. and {Celik}, O. and {Cesarini}, A. and {Ciupik}, L. and {Chow}, Y.~C.~K. and {Cogan}, P. and {Colin}, P. and {Cui}, W. and {Daniel}, M.~K. and {Ergin}, T. and {Falcone}, A.~D. and {Fegan}, S.~J. and {Finley}, J.~P. and {Finnegan}, G. and {Fortin}, P. and {Fortson}, L.~F. and {Furniss}, A. and {Gall}, D. and {Gillanders}, G.~H. and {Grube}, J. and {Guenette}, R. and {Gyuk}, G. and {Hanna}, D. and {Hays}, E. and {Holder}, J. and {Horan}, D. and {Hui}, C.~M. and {Humensky}, T.~B. and {Imran}, A. and {Kaaret}, P. and {Karlsson}, N. and {Kertzman}, M. and {Kieda}, D.~B. and {Konopelko}, A. and {Krawczynski}, H. and {Krennrich}, F. and {Lang}, M.~J. and {LeBohec}, S. and {Lee}, K. and {Maier}, G. and {McCann}, A. and {McCutcheon}, M. and {Moriarty}, P. and {Mukherjee}, R. and {Nagai}, T. and {Niemiec}, J. and {Ong}, R.~A. and {Pandel}, D. and {Perkins}, J.~S. and {Petry}, D. and {Pohl}, M. and {Quinn}, J. and {Ragan}, K. and {Reyes}, L.~C. and {Reynolds}, P.~T. and {Roache}, E. and {Rose}, H.~J. and {Schroedter}, M. and {Sembroski}, G.~H. and {Smith}, A.~W. and {Steele}, D. and {Swordy}, S.~P. and {Toner}, J.~A. and {Vassiliev}, V.~V. and {Wagner}, R. and {Wakely}, S.~P. and {Ward}, J.~E. and {Weekes}, T.~C. and {Weinstein}, A. and {White}, R.~J. and {Williams}, D.~A. and {Wissel}, S.~A. and {Wood}, M. and {Zitzer}, B.},
        title = "{VERITAS Discovery of >200 GeV Gamma-Ray Emission from the Intermediate-Frequency-Peaked BL Lacertae Object W Comae}",
      journal = {\apjl},
         year = 2008,
        month = sep,
       volume = {684},
       number = {2},
        pages = {L73},
          doi = {10.1086/592244},
archivePrefix = {arXiv},
       eprint = {0808.0889},
 primaryClass = {astro-ph},
       adsurl = {https://ui.adsabs.harvard.edu/abs/2008ApJ...684L..73A}
}

@ARTICLE{2011ApJ...726...43A,
       author = {{Abdo}, A.~A. and {Ackermann}, M. and {Ajello}, M. and {Baldini}, L. and {Ballet}, J. and {Barbiellini}, G. and {Bastieri}, D. and {Bechtol}, K. and {Bellazzini}, R. and {Berenji}, B. and {Blandford}, R.~D. and {Bonamente}, E. and {Borgland}, A.~W. and {Bouvier}, A. and {Bregeon}, J. and {Brez}, A. and {Brigida}, M. and {Bruel}, P. and {Buehler}, R. and {Buson}, S. and {Caliandro}, G.~A. and {Cameron}, R.~A. and {Caraveo}, P.~A. and {Carrigan}, S. and {Casandjian}, J.~M. and {Cavazzuti}, E. and {Cecchi}, C. and {{\c{C}}elik}, {\"O}. and {Charles}, E. and {Chekhtman}, A. and {Cheung}, C.~C. and {Chiang}, J. and {Ciprini}, S. and {Claus}, R. and {Cohen-Tanugi}, J. and {Conrad}, J. and {Costamante}, L. and {Cutini}, S. and {Davis}, D.~S. and {Dermer}, C.~D. and {de Palma}, F. and {Digel}, S.~W. and {do Couto e Silva}, E. and {Drell}, P.~S. and {Dubois}, R. and {Dumora}, D. and {Favuzzi}, C. and {Fegan}, S.~J. and {Fortin}, P. and {Frailis}, M. and {Fuhrmann}, L. and {Fukazawa}, Y. and {Funk}, S. and {Fusco}, P. and {Gargano}, F. and {Gasparrini}, D. and {Gehrels}, N. and {Germani}, S. and {Giglietto}, N. and {Giommi}, P. and {Giordano}, F. and {Giroletti}, M. and {Glanzman}, T. and {Godfrey}, G. and {Grenier}, I.~A. and {Grove}, J.~E. and {Guillemot}, L. and {Guiriec}, S. and {Hadasch}, D. and {Hayashida}, M. and {Hays}, E. and {Horan}, D. and {Hughes}, R.~E. and {Itoh}, R. and {J{\'o}hannesson}, G. and {Johnson}, A.~S. and {Johnson}, T.~J. and {Johnson}, W.~N. and {Kamae}, T. and {Katagiri}, H. and {Kataoka}, J. and {Kn{\"o}dlseder}, J. and {Kuss}, M. and {Lande}, J. and {Latronico}, L. and {Lee}, S.-H. and {Longo}, F. and {Loparco}, F. and {Lott}, B. and {Lovellette}, M.~N. and {Lubrano}, P. and {Makeev}, A. and {Mazziotta}, M.~N. and {McEnery}, J.~E. and {Mehault}, J. and {Michelson}, P.~F. and {Mizuno}, T. and {Moiseev}, A.~A. and {Monte}, C. and {Monzani}, M.~E. and {Morselli}, A. and {Moskalenko}, I.~V. and {Murgia}, S. and {Nakamori}, T. and {Naumann-Godo}, M. and {Nestoras}, I. and {Nolan}, P.~L. and {Norris}, J.~P. and {Nuss}, E. and {Ohsugi}, T. and {Okumura}, A. and {Omodei}, N. and {Orlando}, E. and {Ormes}, J.~F. and {Ozaki}, M. and {Paneque}, D. and {Panetta}, J.~H. and {Parent}, D. and {Pelassa}, V. and {Pepe}, M. and {Pesce-Rollins}, M. and {Piron}, F. and {Porter}, T.~A. and {Rain{\`o}}, S. and {Rando}, R. and {Razzano}, M. and {Reimer}, A. and {Reimer}, O. and {Reyes}, L.~C. and {Ripken}, J. and {Ritz}, S. and {Romani}, R.~W. and {Roth}, M. and {Sadrozinski}, H.~F.-W. and {Sanchez}, D. and {Sander}, A. and {Scargle}, J.~D. and {Sgr{\`o}}, C. and {Shaw}, M.~S. and {Smith}, P.~D. and {Spandre}, G. and {Spinelli}, P. and {Strickman}, M.~S. and {Suson}, D.~J. and {Takahashi}, H. and {Tanaka}, T. and {Thayer}, J.~B. and {Thayer}, J.~G. and {Thompson}, D.~J. and {Tibaldo}, L. and {Torres}, D.~F. and {Tosti}, G. and {Tramacere}, A. and {Usher}, T.~L. and {Vandenbroucke}, J. and {Vasileiou}, V. and {Vilchez}, N. and {Vitale}, V. and {Waite}, A.~P. and {Wang}, P. and {Winer}, B.~L. and {Wood}, K.~S. and {Yang}, Z. and {Ylinen}, T. and {Ziegler}, M. and {Acciari}, V.~A. and {Aliu}, E. and {Arlen}, T. and {Aune}, T. and {Beilicke}, M. and {Benbow}, W. and {B{\"o}ttcher}, M. and {Boltuch}, D. and {Bradbury}, S.~M. and {Buckley}, J.~H. and {Bugaev}, V. and {Byrum}, K. and {Cannon}, A. and {Cesarini}, A. and {Christiansen}, J.~L. and {Ciupik}, L. and {Cui}, W. and {de la Calle Perez}, I. and {Dickherber}, R. and {Errando}, M. and {Falcone}, A. and {Finley}, J.~P. and {Finnegan}, G. and {Fortson}, L. and {Furniss}, A. and {Galante}, N. and {Gall}, D. and {Gillanders}, G.~H. and {Godambe}, S. and {Grube}, J. and {Guenette}, R. and {Gyuk}, G. and {Hanna}, D. and {Holder}, J. and {Hui}, C.~M.},
        title = "{Multi-wavelength Observations of the Flaring Gamma-ray Blazar 3C 66A in 2008 October}",
      journal = {\apj},
         year = 2011,
        month = jan,
       volume = {726},
       number = {1},
          eid = {43},
        pages = {43},
          doi = {10.1088/0004-637X/726/1/43},
archivePrefix = {arXiv},
       eprint = {1011.1053},
 primaryClass = {astro-ph.HE},
       adsurl = {https://ui.adsabs.harvard.edu/abs/2011ApJ...726...43A}
}

@ARTICLE{2016MNRAS.461.1862K,
       author = {{Kang}, Shi-Ju and {Zheng}, Yong-Gang and {Wu}, Qingwen and {Chen}, Liang},
        title = "{On the origin of the soft photons of the high-synchrotron-peaked blazar PKS 1424+240}",
      journal = {\mnras},
         year = 2016,
        month = sep,
       volume = {461},
       number = {2},
        pages = {1862-1867},
          doi = {10.1093/mnras/stw1312},
archivePrefix = {arXiv},
       eprint = {1606.00438},
 primaryClass = {astro-ph.HE},
       adsurl = {https://ui.adsabs.harvard.edu/abs/2016MNRAS.461.1862K}
}

@ARTICLE{2025MNRAS.544.1926C,
       author = {{Chen}, Guohai and {Yang}, Wenxin and {Liu}, Yi and {Ho}, Luis C. and {Xiao}, Hubing and {Bachev}, Rumen S. and {Strigachev}, Anton and {Fan}, Junhui},
        title = "{Changing-look behaviour and amplitude-modulated QPO in blazar Ton 599}",
      journal = {\mnras},
         year = 2025,
        month = dec,
       volume = {544},
       number = {2},
        pages = {1926-1938},
          doi = {10.1093/mnras/staf1818},
       adsurl = {https://ui.adsabs.harvard.edu/abs/2025MNRAS.544.1926C}
}

@ARTICLE{2026arXiv260222148B,
       author = {{Ballet}, J. and {Bruel}, P. and {Burnett}, T.~H. and {Lott}, B.},
        title = "{Fermi-LAT 16-year Source List}",
      journal = {arXiv e-prints},
         year = 2026,
        month = feb,
          eid = {arXiv:2602.22148},
        pages = {arXiv:2602.22148},
          doi = {10.48550/arXiv.2602.22148},
archivePrefix = {arXiv},
       eprint = {2602.22148},
 primaryClass = {astro-ph.HE},
       adsurl = {https://ui.adsabs.harvard.edu/abs/2026arXiv260222148B}
}

@ARTICLE{2011PhRvD..83h3507C,
       author = {{Cembranos}, J.~A.~R. and {de La Cruz-Dombriz}, A. and {Dobado}, A. and {Lineros}, R.~A. and {Maroto}, A.~L.},
        title = "{Photon spectra from WIMP annihilation}",
      journal = {\prd},
         year = 2011,
        month = apr,
       volume = {83},
       number = {8},
          eid = {083507},
        pages = {083507},
          doi = {10.1103/PhysRevD.83.083507},
archivePrefix = {arXiv},
       eprint = {1009.4936},
 primaryClass = {hep-ph},
       adsurl = {https://ui.adsabs.harvard.edu/abs/2011PhRvD..83h3507C}
}

@ARTICLE{2022MNRAS.511..938Z,
       author = {{Zeng}, Yuhang and {Yan}, Dahai and {Hu}, Wen and {Wang}, Jiancheng},
        title = "{A spectral hardening in the Fermi-LAT Data of 1ES 0502+675}",
      journal = {\mnras},
         year = 2022,
        month = mar,
       volume = {511},
       number = {1},
        pages = {938-942},
          doi = {10.1093/mnras/stac061},
archivePrefix = {arXiv},
       eprint = {2201.02508},
 primaryClass = {astro-ph.HE},
       adsurl = {https://ui.adsabs.harvard.edu/abs/2022MNRAS.511..938Z}
}

@ARTICLE{2022MNRAS.515.5242D,
       author = {{Dzhatdoev}, T.~A. and {Khalikov}, E.~V. and {Latypova}, V.~S. and {Podlesnyi}, E.~I. and {Vaiman}, I.~A.},
        title = "{Modelling the persistent low-state {\ensuremath{\gamma}}-ray emission of the PKS 1510-089 blazar with electromagnetic cascades initiated in hadronuclear interactions}",
      journal = {\mnras},
         year = 2022,
        month = oct,
       volume = {515},
       number = {4},
        pages = {5242-5250},
          doi = {10.1093/mnras/stac2094},
archivePrefix = {arXiv},
       eprint = {2111.07389},
 primaryClass = {astro-ph.HE},
       adsurl = {https://ui.adsabs.harvard.edu/abs/2022MNRAS.515.5242D}
}

@ARTICLE{2022PhRvD.105j3023C,
       author = {{Cholis}, Ilias and {Zhong}, Yi-Ming and {McDermott}, Samuel D. and {Surdutovich}, Joseph P.},
        title = "{Return of the templates: Revisiting the Galactic Center excess with multimessenger observations}",
      journal = {\prd},
         year = 2022,
        month = may,
       volume = {105},
       number = {10},
          eid = {103023},
        pages = {103023},
          doi = {10.1103/PhysRevD.105.103023},
archivePrefix = {arXiv},
       eprint = {2112.09706},
 primaryClass = {astro-ph.HE},
       adsurl = {https://ui.adsabs.harvard.edu/abs/2022PhRvD.105j3023C}
}

@ARTICLE{2012PhRvD..86h5036T,
       author = {{Tavecchio}, Fabrizio and {Roncadelli}, Marco and {Galanti}, Giorgio and {Bonnoli}, Giacomo},
        title = "{Evidence for an axion-like particle from PKS 1222+216?}",
      journal = {\prd},
         year = 2012,
        month = oct,
       volume = {86},
       number = {8},
          eid = {085036},
        pages = {085036},
          doi = {10.1103/PhysRevD.86.085036},
archivePrefix = {arXiv},
       eprint = {1202.6529},
 primaryClass = {astro-ph.HE},
       adsurl = {https://ui.adsabs.harvard.edu/abs/2012PhRvD..86h5036T}
}

@ARTICLE{2011PhLB..697..412H,
       author = {{Hooper}, Dan and {Goodenough}, Lisa},
        title = "{Dark matter annihilation in the Galactic Center as seen by the Fermi Gamma Ray Space Telescope}",
      journal = {Physics Letters B},
         year = 2011,
        month = mar,
       volume = {697},
       number = {5},
        pages = {412-428},
          doi = {10.1016/j.physletb.2011.02.029},
archivePrefix = {arXiv},
       eprint = {1010.2752},
 primaryClass = {hep-ph},
       adsurl = {https://ui.adsabs.harvard.edu/abs/2011PhLB..697..412H}
}

@ARTICLE{2009arXiv0910.2998G,
       author = {{Goodenough}, Lisa and {Hooper}, Dan},
        title = "{Possible Evidence For Dark Matter Annihilation In The Inner Milky Way From The Fermi Gamma Ray Space Telescope}",
      journal = {arXiv e-prints},
         year = 2009,
        month = oct,
          eid = {arXiv:0910.2998},
        pages = {arXiv:0910.2998},
          doi = {10.48550/arXiv.0910.2998},
archivePrefix = {arXiv},
       eprint = {0910.2998},
 primaryClass = {hep-ph},
       adsurl = {https://ui.adsabs.harvard.edu/abs/2009arXiv0910.2998G}
}

@ARTICLE{2025arXiv251212176W,
       author = {{Wang}, Xiao and {Duan}, Kai-Kai},
        title = "{Constraining the dark matter origin of the halo-like 20 GeV $γ$-ray excess with the AMS-02 antiproton data}",
      journal = {arXiv e-prints},
         year = 2025,
        month = dec,
          eid = {arXiv:2512.12176},
        pages = {arXiv:2512.12176},
          doi = {10.48550/arXiv.2512.12176},
archivePrefix = {arXiv},
       eprint = {2512.12176},
 primaryClass = {astro-ph.HE},
       adsurl = {https://ui.adsabs.harvard.edu/abs/2025arXiv251212176W}
}

@ARTICLE{2025JCAP...11..080T,
       author = {{Totani}, Tomonori},
        title = "{20 GeV halo-like excess of the Galactic diffuse emission and implications for dark matter annihilation}",
      journal = {\jcap},
         year = 2025,
        month = nov,
       volume = {2025},
       number = {11},
          eid = {080},
        pages = {080},
          doi = {10.1088/1475-7516/2025/11/080},
archivePrefix = {arXiv},
       eprint = {2507.07209},
 primaryClass = {astro-ph.HE},
       adsurl = {https://ui.adsabs.harvard.edu/abs/2025JCAP...11..080T}
}

@ARTICLE{2024PhRvD.109l3017Z,
       author = {{Zhong}, Yi-Ming and {Cholis}, Ilias},
        title = "{Robustness of the Galactic Center excess morphology against masking}",
      journal = {\prd},
         year = 2024,
        month = jun,
       volume = {109},
       number = {12},
          eid = {123017},
        pages = {123017},
          doi = {10.1103/PhysRevD.109.123017},
archivePrefix = {arXiv},
       eprint = {2401.02481},
 primaryClass = {astro-ph.HE},
       adsurl = {https://ui.adsabs.harvard.edu/abs/2024PhRvD.109l3017Z}
}

@ARTICLE{2019ARA&A..57..467B,
       author = {{Blandford}, Roger and {Meier}, David and {Readhead}, Anthony},
        title = "{Relativistic Jets from Active Galactic Nuclei}",
      journal = {\araa},
         year = 2019,
        month = aug,
       volume = {57},
        pages = {467-509},
          doi = {10.1146/annurev-astro-081817-051948},
archivePrefix = {arXiv},
       eprint = {1812.06025},
 primaryClass = {astro-ph.HE},
       adsurl = {https://ui.adsabs.harvard.edu/abs/2019ARA&A..57..467B}
}

@ARTICLE{2022ApJS..259...55N,
       author = {{Narendra}, Aditya and {Gibson}, Spencer James and {Dainotti}, Maria Giovanna and {Bogdan}, Malgorzata and {Pollo}, Agnieszka and {Liodakis}, Ioannis and {Poliszczuk}, Artem and {Rinaldi}, Enrico},
        title = "{Predicting the Redshift of Gamma-Ray Loud AGNs Using Supervised Machine Learning. II}",
      journal = {\apjs},
         year = 2022,
        month = apr,
       volume = {259},
       number = {2},
          eid = {55},
        pages = {55},
          doi = {10.3847/1538-4365/ac545a},
archivePrefix = {arXiv},
       eprint = {2201.05374},
 primaryClass = {astro-ph.HE},
       adsurl = {https://ui.adsabs.harvard.edu/abs/2022ApJS..259...55N}
}

@INPROCEEDINGS{2017ICRC...35..824W,
       author = {{Wood}, M. and {Caputo}, R. and {Charles}, E. and {Di Mauro}, M. and {Magill}, J. and {Perkins}, J.~S. and {Fermi-LAT Collaboration}},
        title = "{Fermipy: An open-source Python package for analysis of Fermi-LAT Data}",
    booktitle = {35th International Cosmic Ray Conference (ICRC2017)},
         year = 2017,
       series = {International Cosmic Ray Conference},
       volume = {301},
        month = jul,
          eid = {824},
        pages = {824},
          doi = {10.22323/1.301.0824},
archivePrefix = {arXiv},
       eprint = {1707.09551},
 primaryClass = {astro-ph.IM},
       adsurl = {https://ui.adsabs.harvard.edu/abs/2017ICRC...35..824W}
}

@ARTICLE{2015JCAP...05..011A,
       author = {{Agrawal}, Prateek and {Batell}, Brian and {Fox}, Patrick J. and {Harnik}, Roni},
        title = "{WIMPs at the galactic center}",
      journal = {\jcap},
         year = 2015,
        month = may,
       volume = {2015},
       number = {5},
        pages = {011-011},
          doi = {10.1088/1475-7516/2015/05/011},
archivePrefix = {arXiv},
       eprint = {1411.2592},
 primaryClass = {hep-ph},
       adsurl = {https://ui.adsabs.harvard.edu/abs/2015JCAP...05..011A}
}

@ARTICLE{2026arXiv260400579K,
       author = {{Kang}, Shi-Ju and {Yin}, Yue and {Zheng}, Yong-Gang and {Wu}, Qingwen},
        title = "{A Universal 1.5 GeV Gamma-Ray Line in Active Galactic Nuclei}",
      journal = {arXiv e-prints},
         year = 2026,
        month = apr,
          eid = {arXiv:2604.00579},
        pages = {arXiv:2604.00579},
          doi = {10.48550/arXiv.2604.00579},
archivePrefix = {arXiv},
       eprint = {2604.00579},
 primaryClass = {astro-ph.HE},
       adsurl = {https://ui.adsabs.harvard.edu/abs/2026arXiv260400579K}
}

@ARTICLE{2021JCAP...08..007Z,
       author = {{Zhou}, Jianeng and {Wang}, Zhongxiang and {Huang}, Feng and {Chen}, Liang},
        title = "{A possible blazar spectral irregularity case caused by photon-axionlike-particle oscillations}",
      journal = {\jcap},
         year = 2021,
        month = aug,
       volume = {2021},
       number = {8},
          eid = {007},
        pages = {007},
          doi = {10.1088/1475-7516/2021/08/007},
archivePrefix = {arXiv},
       eprint = {2102.05833},
 primaryClass = {astro-ph.HE},
       adsurl = {https://ui.adsabs.harvard.edu/abs/2021JCAP...08..007Z}
}

@ARTICLE{2025A&A...703A.162D,
       author = {{Dinesh}, A. and {Dominguez}, A. and {Paliya}, V. and {Contreras}, J.~L. and {Buson}, S. and {Ajello}, M.},
        title = "{A systematic search for spectral hardening in blazar flares with the Fermi-Large Area Telescope}",
      journal = {\aap},
         year = 2025,
        month = nov,
       volume = {703},
          eid = {A162},
        pages = {A162},
          doi = {10.1051/0004-6361/202556241},
archivePrefix = {arXiv},
       eprint = {2507.02718},
 primaryClass = {astro-ph.HE},
       adsurl = {https://ui.adsabs.harvard.edu/abs/2025A&A...703A.162D}
}

@ARTICLE{2025ApJ...991L...8P,
       author = {{Paliya}, Vaidehi S. and {B{\"o}ttcher}, M. and {Wani}, Kiran and {Mohammed}, P.~N. Naseef and {Stalin}, C.~S. and {Sahayanathan}, S. and {Saikia}, D.~J. and {Muneer}, S.},
        title = "{The Detection of Teraelectronvolt Radiation from a Flat-spectrum Radio Quasar}",
      journal = {\apjl},
         year = 2025,
        month = sep,
       volume = {991},
       number = {1},
          eid = {L8},
        pages = {L8},
          doi = {10.3847/2041-8213/ae018d},
archivePrefix = {arXiv},
       eprint = {2508.17659},
 primaryClass = {astro-ph.HE},
       adsurl = {https://ui.adsabs.harvard.edu/abs/2025ApJ...991L...8P}
}

@ARTICLE{2025ApJ...988..268L,
       author = {{Li}, Jia-Xuan and {Hu}, Xin-Ke and {Lian}, Ji-Shun and {Yu}, Yu-Wei and {Deng}, Wei and {Liu}, Kuan and {Zhang}, Hai-Ming and {Chen}, Liang and {Zhang}, Jin},
        title = "{Two-component {\ensuremath{\gamma}}-Ray Emission Spectrum and X-Ray Polarization of the Radio Galaxy Pictor A}",
      journal = {\apj},
         year = 2025,
        month = aug,
       volume = {988},
       number = {2},
          eid = {268},
        pages = {268},
          doi = {10.3847/1538-4357/ade808},
archivePrefix = {arXiv},
       eprint = {2407.21277},
 primaryClass = {astro-ph.HE},
       adsurl = {https://ui.adsabs.harvard.edu/abs/2025ApJ...988..268L}
}

@ARTICLE{2019A&A...623A...2A,
       author = {{Ait Benkhali}, Faical and {Chakraborty}, Nachiketa and {Rieger}, Frank M.},
        title = "{Complex gamma-ray behavior of the radio galaxy M 87}",
      journal = {\aap},
         year = 2019,
        month = mar,
       volume = {623},
          eid = {A2},
        pages = {A2},
          doi = {10.1051/0004-6361/201732334},
archivePrefix = {arXiv},
       eprint = {1802.03103},
 primaryClass = {astro-ph.HE},
       adsurl = {https://ui.adsabs.harvard.edu/abs/2019A&A...623A...2A}
}

@ARTICLE{2023arXiv230712546B,
       author = {{Ballet}, J. and {Bruel}, P. and {Burnett}, T.~H. and {Lott}, B. and {The Fermi-LAT collaboration}},
        title = "{Fermi Large Area Telescope Fourth Source Catalog Data Release 4 (4FGL-DR4)}",
      journal = {arXiv e-prints},
         year = 2023,
        month = jul,
          eid = {arXiv:2307.12546},
        pages = {arXiv:2307.12546},
          doi = {10.48550/arXiv.2307.12546},
archivePrefix = {arXiv},
       eprint = {2307.12546},
 primaryClass = {astro-ph.HE},
       adsurl = {https://ui.adsabs.harvard.edu/abs/2023arXiv230712546B}
}

@ARTICLE{2019NewAR..8701541H,
       author = {{Hovatta}, Talvikki and {Lindfors}, Elina},
        title = "{Relativistic Jets of Blazars}",
      journal = {\nar},
         year = 2019,
        month = dec,
       volume = {87},
          eid = {101541},
        pages = {101541},
          doi = {10.1016/j.newar.2020.101541},
archivePrefix = {arXiv},
       eprint = {2003.06322},
 primaryClass = {astro-ph.HE},
       adsurl = {https://ui.adsabs.harvard.edu/abs/2019NewAR..8701541H}
}

@ARTICLE{2019Galax...7...20B,
       author = {{B{\"o}ttcher}, Markus},
        title = "{Progress in Multi-Wavelength and Multi-Messenger Observations of Blazars and Theoretical Challenges}",
      journal = {Galaxies},
         year = 2019,
        month = jan,
       volume = {7},
       number = {1},
          eid = {20},
        pages = {20},
          doi = {10.3390/galaxies7010020},
archivePrefix = {arXiv},
       eprint = {1901.04178},
 primaryClass = {astro-ph.HE},
       adsurl = {https://ui.adsabs.harvard.edu/abs/2019Galax...7...20B}
}

@ARTICLE{2013ApJ...768...54B,
       author = {{B{\"o}ttcher}, M. and {Reimer}, A. and {Sweeney}, K. and {Prakash}, A.},
        title = "{Leptonic and Hadronic Modeling of Fermi-detected Blazars}",
      journal = {\apj},
         year = 2013,
        month = may,
       volume = {768},
       number = {1},
          eid = {54},
        pages = {54},
          doi = {10.1088/0004-637X/768/1/54},
archivePrefix = {arXiv},
       eprint = {1304.0605},
 primaryClass = {astro-ph.HE},
       adsurl = {https://ui.adsabs.harvard.edu/abs/2013ApJ...768...54B}
}

@ARTICLE{2003APh....18..593M,
       author = {{M{\"u}cke}, A. and {Protheroe}, R.~J. and {Engel}, R. and {Rachen}, J.~P. and {Stanev}, T.},
        title = "{BL Lac objects in the synchrotron proton blazar model}",
      journal = {Astroparticle Physics},
         year = 2003,
        month = mar,
       volume = {18},
       number = {6},
        pages = {593-613},
          doi = {10.1016/S0927-6505(02)00185-8},
archivePrefix = {arXiv},
       eprint = {astro-ph/0206164},
 primaryClass = {astro-ph},
       adsurl = {https://ui.adsabs.harvard.edu/abs/2003APh....18..593M}
}

@ARTICLE{1994ApJ...421..153S,
       author = {{Sikora}, Marek and {Begelman}, Mitchell C. and {Rees}, Martin J.},
        title = "{Comptonization of Diffuse Ambient Radiation by a Relativistic Jet: The Source of Gamma Rays from Blazars?}",
      journal = {\apj},
         year = 1994,
        month = jan,
       volume = {421},
        pages = {153},
          doi = {10.1086/173633},
       adsurl = {https://ui.adsabs.harvard.edu/abs/1994ApJ...421..153S}
}
\bibliographystyle{aasjournalv7}

\end{document}